\def \mathbi#1{\textbf{\em #1}}
\documentclass[3p,twocolumn]{elsarticle}
\usepackage{graphicx}
\usepackage{dcolumn}
\usepackage{pifont}
\usepackage{bm}
\usepackage{multirow}
\usepackage{amsmath}

\usepackage{float}
\usepackage{txfonts}

\journal{arXiv}

\begin{document}
\title{Ensemble in phase space: statistical formalism of quantum mechanics}

\author[Jong]{Chol Jong}
\author[Jong]{Byong-Il Ri}
\author[Yu]{Gwang-Dong Yu}
\author[Kim]{Song-Guk Kim}
\author[Jong]{Son-Il Jo}
\author[Jong]{Shin-Hyok Jon}
\author[Jong]{Namchol Choe\corref{cor}}
\ead{cnc103@mail.dlut.edu.cn}

\cortext[cor]{Corresponding author}

\address[Jong]{Department of Theoretical Physics, Faculty of Physics, Kim Chaek University of Technology, Kyogu-60, Central District, Pyongyang,  DPR Korea}
\address[Yu]{Department of Theoretical Physics, Faculty of Physics,  Kim Il Sung University, Ryongnam-Dong, Ryomyong Street, Pyongyang,  DPR Korea}
\address[Kim]{Department of Theoretical Physics, Faculty of Physics,  University of Science, Gwahak-54, Unjong District, Pyongyang, DPR Korea}

\begin{abstract}
We present an alternative formalism of quantum mechanics tailored to  statistical ensemble in phase space. 
The purpose of our work is to show that it is possible to establish an alternative autonomous formalism of quantum mechanics in phase space using statistical methodology.
The adopted perspective leads to obtaining within the framework of its theory the master quantum-mechanical equation without recourse to the other formulations of quantum mechanics, and gives the idea for operators pertaining to dynamical quantities. 
The derivation of this equation starts with the ensemble in phase space and, as a result, reproduces Liouville's theorem and virial theorem for quantum mechanics.
We have explained with the help of this equation the structure of quantum mechanics in phase space and the approximation to the Schr\"{o}dinger equation. 
Furthermore, we have shown that this formalism provides reasonable results of quantization by dealing with some simple cases such as the quantization of harmonic oscillation, the two-slit interference and the uncertainty relation, which confirm the validity of this formalism.
In particular, we have demonstrated that this formalism can easily give the relativistic wave equation without treating the problem of linearizing the Hamiltonian operator by making the most of the point that the master equation is a first-order partial differential equation with respect to time, position and momentum variables, and makes use of the phase velocity.
The ultimate outcome this formalism produces is that primary and general matters of quantum mechanics can be studied reasonably within the framework of statistical mechanics.
\end{abstract}


\maketitle

\section{Introduction}\label{intro}

One of the key questions of quantum mechanics is whether quantum mechanics is able to be established in phase space.
Different opinions about this question gave rise to distinguished formalisms of quantum mechanics.
It is known that there exist three self-standing formulations for quantum mechanics, which involve the conventional Hilbert space, path integral and phase space formalism \cite{Wigner,Groenewold,Zachos,Feynman2,Neuman}. 

The  standard  formalism of  quantum mechanics, i.e. the Hilbert-space formalism  as  developed by Heisenberg, Dirac and von Neumann yielded the successful mathematical  framework for describing the  microworld of atoms and  subatomic objects. This formalism introduces Hermitian operators   to be  able to replace the phase  space  functions of  classical mechanics by mathematical representations in configuration space. 

On  the other hand, since the advent of quantum mechanics, some attempts have been made to modify the standard interpretations and  mathematical formalism of quantum mechanics or to replace them by any other theories.
From the point of view of interpretation, the causal theory of quantum mechanics should be noted. 
The causal theory of quantum mechanics aims to clarify the dynamical causes of quantum-mechanical movements. 
It furnishes the methods of analysis and interpretation for solving quantum dynamical problems, reproducing the concepts of classical mechanics even for quantum mechanics \cite{Goldstein,Goldstein2,Home,Riggs,Wyatt}. 
De Broglie, Madelung, Bohm and others representative of Bohmian mechanics, and Groenewold, Moyal, Takabayasi and others representative of quantum mechanics in phase space (abbreviated as QMPS) had established the foundations of the causal theory of quantum mechanics \cite{Groenewold,Moyal,Takabayasi1,Bohm1,Bohm2,Madelung}. 
Unlike the standard theory of  quantum mechanics, it indispensably adopts the concept of trajectory available even for quantum-mechanical particles \cite{Sbitnev}. 
Of course, the trajectory at issue has a definite probability and is by no means the same as the counterpart in classical mechanics. 

Quantum mechanics with trajectories bifurcates according to whether it makes use of the quantum distribution function defined in phase space or it utilizes the wave function represented in configuration space. 
The hydrodynamic equation of quantum mechanics obtained by Madelung reveals the dynamical characteristics of quantum mechanics \cite{Madelung,Sanz1,Hughes}.  This equation can be readily derived by inserting into the Schr\"{o}dinger equation the wave function in the polar form represented by the action. 

This formalism was greatly extended by D. Bohm, who showed how the quantal effect originating from quantum potential affects the movements of microparticles \cite{Bohm1}.
Bohmian mechanics works on the basis of three fundamental assumptions, viz. the assumption about wave field satisfying the Schr\"{o}dinger equation, form of wave function and statistical ensemble \cite{Bohm1,Bohm2}.

Bohmian mechanics claims that all the results obtained by the standard theory of quantum mechanics can be derived on the basis of these assumptions and can be deeply interpreted with a natural epistemology.
It definitely determines the momentum of a particle via the wave function.
As a result, the basic system of equations in Bohmian mechanics should be considered as framing a phase-space ensemble, according to the following diagram:
\begin{equation*}\label{eq:spec1}
\mathbi{q} ~ \rightarrow ~ \psi \left( \mathbi{q} \right) ~ \rightarrow ~ \hat{\mathbi{p}}\psi \left( \mathbi{q} \right)~\rightarrow~\mathbi{p} ~.
\end{equation*}	

The curvature obtained by integrating the equation for velocity determines the trajectories of particles. 
Essentially, Bohmian mechanics is a deterministic quantum theory for point-like particles since it is based on the concepts pertaining to  ensemble and trajectory. 
This formalism employs the Schr\"{o}dinger equation, but provides alternative conceptions and methods of dynamical interpretation of quantum mechanics  distinguished from those of the standard theory of quantum mechanics.

The novel quantum-trajectory method is finding a broad range of application to such dynamical problems as mixed quantum-classical dynamics, density matrix evolution in dissipative systems and electronic non-adiabatic dynamics \cite{Wyatt,durr,Meier,Ghosh,Chou}.

From the point of view of mathematical formalism, it should be noted that as an autonomous formalism of quantum mechanics, QMPS has received considerable attention \cite{Zachos,Martens}. 
This formalism of quantum mechanics is distinguished from the conventional operator theory in Hilbert space. 
For this self-standing formulation, there is no need to choose sides between position and momentum space \cite{Zachos}.
It is based on Wigner's quasi-distribution  function and  Weyl's correspondence between  quantum-mechanical  operators in Hilbert  space  and  ordinary  complex-valued functions in phase  space \cite{Weyl,Baker,Takabayasi2}.  
The structure of  Wigner's quasi-distribution  function  was  fully interpreted  by Groenewold and  Moyal, and  insights into interpretation and an appreciation of its conceptual autonomy were developed by Takabayasi, Baker, Fairlie and others \cite{Groenewold,Moyal,Takabayasi1,Baker,Fairlie}. 
It works in full phase space, satisfying the uncertainty principle \cite{Takabayasi1}, and provides real insights into several problems in quantum mechanics including quantum transport process and transition to classical statistical mechanics.

Mapping from a wave function to a distribution function in phase space is not unique.
An alternative form of distribution function in phase space \cite{Wyatt} can be easily given provided it is considered that the wave function in momentum space, $\phi\left(\mathbi{p}\right)$ is obtained by means of the Fourier transformation of the wave function in position space, $\psi\left(\mathbi{q}\right)$, and the wave function in phase space is the multiplication of the two wave function, i.e. $\psi\left(\mathbi{q}\right)$ and $\phi\left(\mathbi{p}\right)$.
The equation of motion for the probability density is obtained by differentiating both sides of the definition expression for probability density with respect to time, and then by taking into consideration the Schr\"{o}dinger equation \cite{Zachos}.
Thus, QMPS starts with the distribution function in phase space and develops its theory \cite{Wyatt,Fairlie}. 

There have been several investigations for systematic developments of theories and remarkable contributions to applications in this field \cite{Neuman,Fairlie,Royer,Dahl,Curtright}.
Iafrate \textit{et al} and Gardner developed the equation of motion for the momentum moments of the Wigner function \cite{Iafrate,Gardner}. 
Gasser and Markowich studied semi-classical and classical limits of quantum transport equations derived from the momentum moments of the Wigner function \cite{Gasser}.  
Moreover, Muga carried out an investigation on the connection between moments and quantum phase space distributions \cite{Muga}. 

This formalism of quantum mechanics is useful to describe quantum transport processes, which are of importance in several fields including quantum optics, condensed matter physics, the semi-classical limits of mesoscopic systems, and the transition to classical statistical mechanics \cite{Gasser,Schleich}. 
In this regard, numerous investigations for extending applications of quantum mechanics in phase space have received a growing interest. 
\cite{Burghardt1,Burghardt2,Burghardt3,Maddox,Bittner}.  

On the other hand, the tomographic formulation of quantum mechanics \cite{Nicola,Monsa,Arkhipov} has received considerable attention in recent years. For this approach, the dynamical variables of the theory are a set of probability distributions, which has truly classical-like characteristics such as non-negative, normalized and, in principle, all measurable.
Symplectic tomograms can be obtained by an integral transform of $\psi \left( x \right )$ describing a quantum normalized state.
These tomograms are used to reconstruct the density matrix as a complex function by means of inverse transform. 
All other characteristics, such as the Wigner function, can be also expressed in terms of the symplectic tomogram. Thus, the theory of quantum tomography treats the mapping of the quantum states in position space to ones in phase space. 

An attracting question of quantum mechanics in phase space is whether it is possible to establish the formalism as a self-standing.
In this connection, it is Moyal's equation that is extremely unusual among several versions of QMPS \cite{Moyal}.
Moyal's equation is obtained by introducing probability density in terms of the statistical ensemble in phase space, and by taking into consideration the requirement that the equation of movement represented by means of the big star operator \cite{Moyal} should coincide with Liouville's theorem within classical limits. 
It is possible to completely interpret a series of quantum-mechanical problems with the help of Moyal's equation \cite{Zachos,Curtright2}.  

The resolution of the problem of whether quantum mechanics in phase space is able to become another autonomous formalism may lead to the formation of more general formalism of quantum mechanics.
In this connection, it is noticeable that Moyal's method does not employ the Schr\"{o}dinger equation and assumes an autonomous formalism of quantum mechanics  \cite{Moyal,Curtright2}. 
Moyal's equation shows that there may be other ways capable of describing quantal phenomena without recourse to the Schr\"{o}dinger equation. 

Meanwhile, It is generally seen that  the idea of quantum mechanics with trajectories is identified with that of the path integral formulation which was conceived by Dirac and then was constructed by Feynman \cite{Feynman}.

This status of quantum mechanics implies that quantum mechanics still has not been  satisfactorily framed in the aspect of formalism and for this reason there may be distinct self-standing formalisms in future.
A newly established formalism should illuminate inconsistent aspects of the present theories of quantum mechanics, thus giving a better foundation and interpretation to the quantum theory.
The theory of quantum mechanics in phase space offers the prospect of improving the formalism, since it works in phase space representing complete information on dynamical movement.
With such an understanding, we aim to present an alternative formalism of quantization in terms of statistical ensemble  in phase space demonstrating the probabilistic and mechanical structure of quantum mechanics, or wave-particle properties of quantum systems. 
To be an autonomous formalism, it must yield its master equation independently of  other formalisms.
To achieve the goal to formulate an autonomous formalism, our work starts with the statistical ensemble in phase space. 
Our work shows that quantum mechanics can accept the phase-space formalism as providing a generalized theory of quantum mechanics in a consistent manner.
In fact, manipulating in phase space, our methodology makes a comprehensible, natural inference from the probability wave to produce an alternative master quantum-mechanical equation without recourse to the other formulations of quantization, thereby explaining a series of quantization problems. 

In conclusion, we have grounds to conclude that  there is a possibility of providing a new route to quantum mechanics.

\section{Strategy for obtaining master equation of QMPS}

An autonomous formalism indispensably requires its independent master equation.
The Schr\"{o}dinger equation, which furnishes the quantum-mechanical state function for microparticles, i.e. the wave function, embodies the fundamental concepts and methodologies of quantum mechanics. 

The Schr\"{o}dinger equation can be guessed at in several ways  \cite{Greiner,Landau}.
Fermi showed in his book ``Notes on Quantum Mechanics" that the ordinary wave equation can be transformed into the Schr\"{o}dinger equation in view of de Broglie's relation.
Without some assumptions, it is impossible to achieve the goal for obtaining the Schr\"{o}dinger equation.
 
After the discovery of the Schr\"{o}dinger equation, the Klein-Gordon equation and Dirac's equation were proposed as relativistic wave equations. 
These equations cannot be derived from a certain generalized basic equation of quantum mechanics and should be made with the help of the operators corresponding to physical quantities, inferred from the Schr\"{o}dinger equation.

Consequently, the Schr\"{o}dinger equation amounts to the basic premise for quantum mechanics in all respects and the operators, rather than the equation, have general meaning. 
Therefore it is hardly too much to say that in practice quantum mechanics is based on the Schr\"{o}dinger equation. 
The powerful argument demonstrating the validity of the Schr\"{o}dinger equation is the fact that the results obtained by solving this equation for quantum objects are in good agreement with a wide range of experiments.

For this reason, there have not been so many disputes as to whether or not the Schr\"{o}dinger equation is exact. So several schools, which take different conceptual and philosophical interpretations in quantum mechanics, commonly adopted the Schr\"{o}dinger equation and have been developing their theories using it.
By and large, the Schr\"{o}dinger equation is exact.

However it is necessary to note that there is no need to disregard the possibilities of adopting any other formalism independent of the Schr\"{o}dinger equation, since it could elucidate incomplete aspects of quantum mechanics and resolve some knotty problems.
In this sense we seek a novel master equation of quantum mechanics in phase space inferred from the probability wave.

Our strategy for obtaining an alternative master equation of QMPS is based on the conceptions of the wave field and statistical ensemble in phase space. 
Our formalism works in phase space, based on the views on the statistical structure of quantum mechanics. 
Without using the Schr\"{o}dinger equation, we obtain a new master equation in phase space from the picture of statistical ensemble representing the wave field.

We aim to develop an alternative formalism of QMPS by rationally inheriting the theoretical successes achieved by both the standard  and the causal theory of quantum mechanics.
The proposed master equation of QMPS is represented in phase space and contains both the probability continuity equation and the dynamical relation of particles, and in addition explains an idea of why to introduce operators. 
\section{Wave function defined in phase space and master equation of QMPS}

\subsection{Basic assumptions}

This formalism uses the wave function and probability density in phase space.
Starting with statistical ensemble in phase space, it yields its autonomous master equation in phase space.

To frame another formalism of quantum mechanics in phase space, we form a set of assumptions as follows.

\textbf{Assumption I:} \textit{The phase space specifies states of  microparticles and the uncertainty of microparticles is expressed as the statistical ensemble in phase space.}

Essentially, the statistical ensemble demonstrates the wave field as a physical reality, thus exhibiting the dynamical causality in the microworld.
In this formalism, both position and momentum are basic variables of quantum-mechanical state functions. 
Introducing both positions and momenta as dynamical variables enables us to make quantum mechanics resemble  statistical mechanics in respect of methodologies, thereby developing perspectives on quantum mechanics.

The problem is whether the position and momentum variables of a particles can be utilized together to represent states of microparticles. 
Taking both position and momentum variable as basic variables of a quantum-mechanical state does not violate the uncertainty principle.
In quantum mechanics, positions of particles are used to indicate the probability of  finding particles in a volume element of configuration space. 
From the point of view of statistical interpretation, it is obvious that the Schr\"{o}dinger equation deals with ensembles in configuration space for positions of particles. 

It is conceivable that a distribution in position space naturally yields a definite distribution in momentum space, provided that a position corresponds to a momentum.
On the one hand, it is possible to determine the wave function in momentum space by performing the Fourier transform of a wave function in position space.
Obviously, it denies one-to-one correspondence of position and momentum.
On the other hand, it  is necessary to deliberate on the fact that the application of the momentum operator to a wave function definitely determines particles' momenta.
In doing so, a momentum operator makes a position uniquely correspond  to a definite momentum through a wave function.
This relation can be schematically represented as
\begin{equation}\label{eq:sp1}
\mathbi{q}~ \rightarrow ~ \psi \left( \mathbi{q} \right) ~\rightarrow~ \hat{\mathbi{p}_i} \psi \left( \mathbi{q} \right) ~\rightarrow ~\mathbi{p}~,
\end{equation}
where $\hat{\mathbi{p}_i}$ is the momentum operator for the $i$th particle, $\mathbi{q}$ the whole of coordinates of particles and $\mathbi{p}$ the whole of momentum components of particles.
As a consequence, $ \mathbi{q}~\rightarrow~ \mathbi{p}$.
Thus, we are in a position to imagine a definite set of a position and the corresponding momentum.  
In this context, an ensemble in position space definitely corresponds to that in momentum space.
In the end, it is concluded that joining these spaces gives a phase space.

The point is how these two distributions correlate.
Of course, this relation may be rigorously proved, thereby showing that the uncertainty relation is derived by treating the statistical ensemble. 

The quantal ensemble essentially differs from the classical one. 
The classical ensemble is given by physical objects that obey the very same physical law, but are distinguished only by initial conditions with randomness. 
For a classical ensemble, the correlation between phase trajectories does not exist. 
For this reason, the probabilistic characteristics are exhibited due to the randomness of initial conditions. 
On the other hand, the quantal ensemble displays probabilistic characteristic which characterizes the wave field expressed as the correlation between phase trajectories, i.e. self-interference. 
Then the phase points in the quantal ensemble are governed by the correlation according to de Broglie's relation, i.e. the wave-like property. 
Therefore the quantal ensemble should assume both the dynamical laws for particles and the wave-like property characterizing the wave field. 

In the first place, we define the density of phase points given by a statistical ensemble in phase space which represents the probability of finding particles in the volume element centered on a phase point $\left(\mathbi{q}, \mathbi{p}\right)$ as
\begin{equation}\label{eq:14}
\rho=\rho\left(\mathbi{q},\mathbi{p},t \right)~.
\end{equation}
In fact, such a definition is natural in statistical mechanics. 

In the second place, we assume that the probability density is given by the square of modulus of the wave function in phase space.

Thus, we have
\begin{equation}\label{eq:15}
\rho=\psi^\ast \left(\mathbi{q},\mathbi{p},t \right) \psi \left(\mathbi{q},\mathbi{p},t \right) ~.  
\end{equation}

According to this definition, the requirement is implemented that the probability density should always be real-valued and positive, and the wave function should be tailored so as to describe the coherence of the wave field. 

It is necessary to note that early in 1950s Wigner, Moyal, Groenewold and Takabayasi constructed QMPS on the basis of the probability density defined in phase space. 
In spite of all detailed differences from the other phase-space formulations, our formalism is akin to them in that it starts with the probability density defined in phase space. 

Obviously, the wave function should satisfy the following normalization condition:
\begin{equation}\label{eq:16}
\int\int \psi^\ast \left(\mathbi{q},\mathbi{p},t \right) \psi \left(\mathbi{q},\mathbi{p},t \right) d\mathbi{q} d\mathbi{p}=1 ~.
\end{equation}

By definition, the probability density in configuration space is determined by
\begin{equation*}\label{eq:17}
\rho_q(\mathbi{q})=\int \rho(\mathbi{q},\mathbi{p})d\mathbi{p}~, 
\end{equation*}
again, the probability density in momentum space, by
\begin{equation*}\label{eq:18}
\rho_p(\mathbi{p})=\int \rho(\mathbi{q},\mathbi{p})d\mathbi{q} ~.
\end{equation*}

Furthermore, the mean value of a physical quantity is defined as 
\begin{align}\label{eq:19}
\bar{F}= \int\int \psi^\ast(\mathbi{q},\mathbi{p}) F(\mathbi{q},\mathbi{p})\psi(\mathbi{q},\mathbi{p})  d \mathbi{q} d \mathbi{p} ~.
\end{align}

\textbf{Assumption II:} \textit{De Broglie's relation specifies the correlation between particles and the wave field which encompasses the particles and is inseparable from them.}

In essence, it is merely the introduction of de Broglie's relation to this formalism, but emphasis is placed on the wave field.
Note that the Schr\"{o}dinger equation is obtained by finding the operator equation suggesting the energy relation of classical mechanics, starting with the wave function of a free particle. 
In fact, without de Broglie's relation it is impossible to imagine the wave function of a free particle. 
Evidently, de Broglie's relation characterizes the wave field which yields a statistical ensemble.

\textbf{Assumption III:} \textit{A wave function is expressed as the product of an amplitude part and a phase one described by the action.}

Thus for a many-particle system, we writes wave functions as 
\begin{equation}\label{eq:20}
\psi(\mathbi{q},\mathbi{p},t)=\psi_0(\mathbi{q},\mathbi{p},t) \exp (iS/\hbar)~,    
\end{equation}
where $\psi_0$  is a real-valued function and $S$, the action as
\begin{equation}\label{eq:21}
S=\int \mathbi{p}~ \mathrm{d}\mathbi{q} -\int H \mathrm{d}t ~,
\end{equation}
where \mathbi{q} designates the whole of coordinates, \mathbi{p} that of momentum components of the system under consideration.
The action which reflects the ensemble of trajectories of a given system can be considered as a characteristic integral in the context of the Poincar\'{e}-Cartan integral invariant. 

This assumption implies that the wave field has the phase determined by the action, and $h$ is the quantum of the action.
Comparing de Broglie' relation with Eq. \ref{eq:21} naturally leads to grasping sound meanings of the dualism.
Also, this assumption is not regarded to be new, since such a form of wave function has already been used in the preceding formalisms \cite{Poirier}.
It is necessary to recall the fact that the the Schr\"{o}dinger equation was obtained,  implicitly employing this assumption.
In fact, for the  Schr\"{o}dinger equation the phase part of the wave function assumed for a free particle agrees with this assumption. 
The subject relevant to the phase part will be concretely discussed in Summary and Discussion.
These three assumptions about the wave field serve as the basis for establishing  the present formalism of quantum mechanics in phase space.

\subsection{\label{subsec:mas_eq}Master equation of quantum mechanics in phase space}

To obtain the master equation of quantum mechanics in phase space,  we start from statistical ensemble in phase  space.

According to Liouville's theorem, we have the following equation of motion for the density of phase points:
\begin{equation}\label{eq:22_15}
\frac{\mathrm{d} \rho}{\mathrm{d}t}=\frac{\partial \rho}{\partial t}+\sum_{i=1}^{f}\left[ \dot{q}_i \frac{\partial \rho}{\partial q_i}+\dot{p}_i \frac{\partial \rho}{\partial p_i} \right]=0 ~, 
\end{equation}
where $f$ is the number of degrees of freedom, $q_i$ and $p_i$ are the generalized coordinates and momenta.

Therefore we have
\begin{equation}\label{eq:23}
\boxed{\frac{\partial \rho}{\partial t}=-\sum_{i=1}^{f} \left[ \dot{q}_i \frac{\partial \rho}{\partial q_i}+\dot{p}_i \frac{\partial \rho}{\partial p_i} \right] }~. 
\end{equation}

The above equation is a wave equation in the form of first-order partial differential equation. 
Since the probability density behaves like a wave, the derivative of position with respect to time and the derivative of momentum with respect to time in the above equation are considered as quantities pertaining to a wave, and come to have no longer the meanings of particle-like quantities. 
When taking into account that $\frac{\mathrm{d} \rho}{\mathrm{d} t}$ is constant on the wave front of $\rho$, we can easily understand Eq. \ref{eq:22_15}.

The quantal ensemble exhibits wave-like properties. 
Therefore the quantal causality of microparticles should be taken into account by replacing wave-like quantities by the corresponding particle-like ones. 
On the other hand, it is impossible to consider a change in the density, following a phase point determined by particles in such a way as done when deducing Liouville's theorem. 
The meaningful velocity for such a consideration for a wave is the phase velocity. 
Accordingly, it is necessary to consider a change in probability density in a system moving at the phase velocity of probability wave, since it propagates at the phase velocity. 
In this case, we should express the phase velocities as the corresponding group velocities with the help of the relation between phase and group velocity. 
De Broglie's relation gives the following relation between phase and group velocity:
\begin{equation*}\label{eq:24}
v_{phase}=\frac{\omega}{k}=\frac{E}{p} ~,
\end{equation*}
\begin{equation}\label{eq:25}
v_{phase} \cdot v_{group}=\frac{E}{m} ~.
\end{equation}

Considering only the kinetic energy of a particle as the energy fulfilling de Broglie's relation leads to the general relation between phase and group velocity,
\begin{equation}\label{eq:26}
v_{phase}=\frac{v_{group}}{2} ~.
\end{equation}

It is obvious that for non-relativistic case, as the relation between phase and group velocity is linear, so  the relation between phase and group acceleration. 

Then we have
\begin{equation}\label{eq:28}
v_{phase}=\frac{v_{group}}{2},\:\:\:\frac{\mathrm{d}v_{phase}}{\mathrm{d}t}=\frac{1}{2}\frac{\mathrm{d}v_{group}}{\mathrm{d}t} ~.
\end{equation}

Note that for a particle in a potential field, de Broglie's relation should be extended. 
For a particle in a potential field the phase velocity should be written as 
\begin{equation}\label{eq:29}
v_{phase}=\frac{{E-U}}{\sqrt{2 m(E-U)}} ~.
\end{equation}

Unless such a generalized concept is accepted, the de Broglie relation does not assume generality and it is inevitable that the quantum theory will be faced with many of intractable problems. 
Holland already discussed this problem and gave the same idea  \cite{Holland}. 
This matter will be discussed in detail later.
It is concluded that only kinetic energy is related to the wave-like characteristics. 
For the Schr\"{o}dinger equation, use was already made of such a relation to make the equation for free particles available even for the case of the particles in a potential field. 

Our consideration starts with the probability continuity equation. 
For a probability wave, the probability continuity equation should hold since the probability is conserved.

Then we have
\begin{equation}\label{eq:30}
\frac{\partial \rho}{\partial t}+ \nabla \left( \rho \mathbi{V}_{phase} \right)=0~,
\end{equation}
and in a further step,
\begin{equation*}\label{eq:31}
\frac{\partial \rho}{\partial t}+ \rho \nabla \mathbi{V}_{phase}+\mathbi{V}_{phase} \nabla \rho=0~.
\end{equation*}

Taking into consideration the linear relation between phase and group velocity, we get the following results:
\begin{equation}\label{eq:32}
\nabla \mathbi{V}_{phase} =0~.
\end{equation}

Consequently, it is obvious that the probability behaves like an incompressible fluid. 
Hence the probability continuity equation is represented as 
\begin{equation}\label{eq:33}
\frac{\partial \rho}{\partial t}+\mathbi{V}_{phase} \nabla \rho=0~.
\end{equation}

Consequently, for a quantal ensemble behaving like a wave, the probability continuity equation, 
\begin{equation}\label{eq:33a}
\frac{\mathrm{d} \rho}{\mathrm{d} t}=0 
\end{equation}
holds, which we may as well refer to as Liouville's theorem for quantal ensemble.
We adopt Eq. \ref{eq:33a} as one valid for both relativistic and non-relativistic case.
After inserting Eq. \ref{eq:15} into Eq. \ref{eq:33} to obtain the wave equation, we go through the following steps:
\begin{eqnarray*}\label{eq:34}
\frac{\partial \left( \psi^\ast \psi\right)}{\partial t}+\sum_{i=1}^{f}\left[ \dot{q}_i \frac{\partial \left( \psi^\ast \psi\right)}{\partial q_i}+\dot{p}_i \frac{\partial \left( \psi^\ast \psi\right)}{\partial p_i} \right]=0~,
\end{eqnarray*}
\begin{align}\label{eq:35}
&\psi\frac{\partial \psi^\ast }{\partial t}+\sum_{i=1}^{f}\psi \left[ \dot{q}_i \frac{\partial \psi^\ast}{\partial q_i}+\dot{p}_i \frac{\partial \psi^\ast}{\partial p_i} \right]\nonumber \nonumber \\
&+\psi^\ast\frac{\partial \psi}{\partial t}+\sum_{i=1}^{f}\psi^\ast \left[ \dot{q}_i \frac{\partial \psi}{\partial q_i}+\dot{p}_i \frac{\partial \psi}{\partial p_i} \right]=0~,
\end{align}
\begin{align}\label{eq:36_31}
\psi^\ast\frac{\partial \psi }{\partial t}+\sum_{i=1}^{f}\psi^\ast\left[ \dot{q}_i \frac{\partial \psi}{\partial q_i}+\dot{p}_i \frac{\partial \psi}{\partial p_i} \right]\nonumber \\
+\textit{complex conjugate}=0 ~,
\end{align}
where $\dot{q}_i$ and $\dot{p}_i$  are velocity components in phase space.

Dividing both sides of the above equation by $\psi^\ast \psi$ gives 
\begin{align}\label{eq:37}
\frac{1}{\psi}\frac{\partial \psi }{\partial t}+\frac{1}{\psi}\sum_{i=1}^{f} \left[ \dot{q}_i \frac{\partial \psi}{\partial q_i}+\dot{p}_i \frac{\partial \psi}{\partial p_i} \right]\nonumber \\
+\textit{complex conjugate}=0~.
\end{align}

The left side of Eq. \ref{eq:37} is real-valued. 
From this, it follows that only the real part of the expression,
\begin{equation*}\label{eq:38}
\frac{1}{\psi}\frac{\partial \psi }{\partial t}+\frac{1}{\psi}\sum_{i=1}^{f} \left[ \dot{q}_i \frac{\partial \psi}{\partial q_i}+\dot{p}_i \frac{\partial \psi}{\partial p_i} \right] 
\end{equation*}
is significant.

Finally, we get the wave equation,
\begin{equation}\label{eq:39}
\boxed{\frac{\partial \psi }{\partial t}=-\sum_{i=1}^{f}\left[ \dot{q}_i \frac{\partial \psi}{\partial q_i}+\dot{p}_i \frac{\partial \psi}{\partial p_i} \right]} ~.
\end{equation}

Generally, the solution to Eq. \ref{eq:39}, $\psi$ is a complex-valued function.
Obviously, Eq. \ref{eq:39} is the wave equation in phase space that is represented as a first-order differential equation. 
For this reason, $\psi$ can be referred to as the wave function.

In the next place, we substitute particle-like quantities for wave-like quantities. 
$\dot{q}_i$ issuing from the wave function must be considered as the phase velocity of a wave, since Eq. \ref{eq:39} is a wave equation. 
Now, it is possible to introduce instead of the phase velocities in Eq. \ref{eq:39} the group velocities represented by means of the Hamiltonian function,
\begin{equation*}\label{eq:40}
\dot{q}_{group}=\frac{\partial H}{\partial p}~,
\end{equation*}
since using de Broglie's relation, we can express the phase velocities  by means of the group velocities.
For $\dot{p}_i$, we can also do one and the same. 
It is obvious that $\dot{q}_i$ is the phase velocity in configuration space, while $\dot{p}_i$ is the phase velocity in momentum space.
With the help of the relationship between phase and group velocity, we get the wave equation written by means of group velocities, 
\begin{equation}\label{eq:41}
\boxed{\frac{\partial \psi }{\partial t}=-\frac{1}{2}\sum_{i=1}^{f}\left[ \frac{\partial H}{\partial p_i} \frac{\partial \psi}{\partial q_i}-\frac{\partial H}{\partial q_i} \frac{\partial \psi}{\partial p_i} \right]}~.
\end{equation}

Next, we manipulate and scrutinize the above equation to conceive the notion of operator.
For the purpose of searching for operators, multiplying both sides of this equation by $i\hbar$, we have in the purely heuristic form
\begin{equation}\label{eq:42_37}
\boxed{i\hbar\frac{\partial \psi }{\partial t}=-\frac{i\hbar}{2}\sum_{i=1}^{f} \left[ \frac{\partial H}{\partial p_i} \frac{\partial \psi}{\partial q_i}-\frac{\partial H}{\partial q_i} \frac{\partial \psi}{\partial p_i} \right]}~.
\end{equation}

Of course, such an operation does not endow the equation with any new meaning from the point of view of mathematics. 
However, it enables us to perceive operators as the tool for deriving dynamical quantities from the wave function. 
The application of a differential operator to wave function should yield a corresponding dynamical quantity.
After due consideration, we arrive at finding the operator relations equal or analogous to ones in the Schr\"{o}dinger equation.

The explanation runs as follows. 
What should be stressed is that the results are due to the assumed form of wave function.
To begin with, we calculate the derivatives of the action in the extended phase space, 
\begin{equation}\label{eq:42-38}
S\left(\mathbi{q}, \mathbi{p}, t\right)=\int_0^{\mathbi{q}} \mathbi{p}~ \mathrm{d}\mathbi{q}' -\int_0^t H \mathrm{d}t'~
\end{equation}
with respect to $q, p, t$.

The results are as follows.
\begin{equation}\label{eq:42-39}
\frac{\partial S\left(\mathbi{q}, \mathbi{p}, t\right)}{\partial q_i} =p_i,
\end{equation}
\begin{equation}\label{eq:42-40}
\frac{\partial S\left(\mathbi{q}, \mathbi{p}, t\right)}{\partial p_i} =q_i,
\end{equation}
\begin{equation}\label{eq:42-41}
\frac{\partial S\left(\mathbi{q}, \mathbi{p}, t\right)}{\partial t} =-H.
\end{equation}
Next, let us review $i\hbar \dfrac{\partial \psi}{\partial t}$.
Recall the following assumed form of wave function:
\begin{equation}\label{eq:43}
\psi \left(\mathbi{q},\mathbi{p},t\right)=\varphi\left(\mathbi{q},\mathbi{p},t\right)\exp\left\{\dfrac{iS\left(\mathbi{q},\mathbi{p},t\right)}{\hbar}\right\} 
\end{equation}
where $\varphi\left(\mathbi{q},\mathbi{p},t\right)$ is a real-valued function.
The application of $i\hbar \dfrac{\partial}{\partial t}$ to the wave function yields
\begin{align}\label{eq:44}
i\hbar \frac{\partial \psi}{\partial t}=i\hbar \frac{i}{\hbar}\frac{\partial S}{\partial t}\:\psi+i\hbar \frac{1}{\varphi}\frac{\partial \varphi}{\partial t}\:\psi
=H \psi+i\hbar \frac{1}{\varphi} \frac{\partial \varphi}{\partial t}\:\psi ~.
\end{align}
From the above expression, we can interpret $i\hbar \dfrac{\partial}{\partial t}$ as the operator relative to total energy, since it makes $H$ be derived from the wave function.

Furthermore, let us examine $-i\hbar \dfrac{\partial \psi}{\partial p_i}$.
The application of this operator to the wave function produces
\begin{align}\label{eq:44-1}
-i\hbar \frac{\partial \psi}{\partial p_i}=-i\hbar \frac{i}{\hbar}\frac{\partial S}{\partial p_i}\:\psi-i\hbar \frac{1}{\varphi}\frac{\partial \varphi}{\partial p_i}\:\psi
=q_i \psi-i\hbar \frac{1}{\varphi} \frac{\partial \varphi}{\partial p_i}\:\psi ~.
\end{align}
Since this operation gives $q_i$, we can regard $-i\hbar \dfrac{\partial}{\partial p_i}$ as the position operator.

Similarly, we have
\begin{align}\label{eq:44-2}
-i\hbar \frac{\partial \psi}{\partial q_i}=-i\hbar \frac{i}{\hbar}\frac{\partial S}{\partial q_i}\:\psi-i\hbar \frac{1}{\varphi}\frac{\partial \varphi}{\partial q_i}\:\psi
=p_i \psi-i\hbar \frac{1}{\varphi} \frac{\partial \varphi}{\partial q_i}\:\psi ~.
\end{align}
As a consequence, $-i\hbar \dfrac{\partial}{\partial q_i}$ is adopted as the momentum operator.
The obtained results naturally bring the idea of operator to us. 
From Eqs. \ref{eq:44}-\ref{eq:44-2}, we can interpret the meaning of the relation between an observable, $L$ and the corresponding operator, $\hat{L}$ as 
\begin{equation}\label{eq:45a}
L=\mathrm{Re} \left( \frac{1}{\psi} \hat{L} \psi \right)~.
\end{equation}
As a matter of fact, this relation naturally comes from the difinition of mean value.
By difinition, the mean value with respect to $\hat{L}$ is written as
\begin{equation}\label{eq:45b}
\bar{L}=\int \psi^\ast \hat{L} \psi \mathrm{d}\tau=\int \psi^\ast \psi \frac{\hat{L} \psi}{\psi} \mathrm{d}\tau=\int \left|\psi\right|^2 \frac{\hat{L} \psi}{\psi} \mathrm{d}\tau~.
\end{equation}
Accordingly, $\mathrm{Re} \left( \dfrac{1}{\psi} \hat{L} \psi \right)$ should be regarded as the observable with respect to operator $\hat{L}$.

By inference, we find the operators corresponding to kinetic energy and potential. 
Altogether, the operators corresponding to fundamental observables are represented as
\begin{subequations}
\begin{gather}
\hat{E}=i\hbar \frac{\partial}{\partial t},\\
\hat{p_i}=-i\hbar \frac{\partial}{\partial q_i},\\
\hat{q}_i=-i\hbar \frac{\partial}{\partial p_i},\\
\hat{U}=-\frac{i\hbar}{2} \sum_{i=1}^{f} -\frac{\partial H}{\partial q_i}\:\frac{\partial}{\partial p_i}= \frac{1}{2}\sum_{i=1}^{f} \dot{p}_i~\hat{q}_i,\\
\hat{T}=-\frac{i\hbar}{2}\sum_{i=1}^{f} \frac{\partial H}{\partial p_i} \frac{\partial }{\partial q_i}=\frac{1}{2} \sum_{i=1}^{f}\dot{q}_i\:\hat{p}_i~.
\end{gather}
\end{subequations}

The successive three operators correspond to energy, momentum and position respectively, which become basic dynamical quantities. 
The fourth operator should be considered as the potential energy operator since it corresponds to a potential energy function. 
This operator suggests nothing other than the $\mathbi{virial theorem}$ of statistical mechanics.
Thus, we can arrive at the important conclusion that in quantum mechanics the potential energy should be represented as the virial of the system under consideration.

Meanwhile, the fifth operator should be considered as the kinetic energy operator, since it corresponds to kinetic energy. 

The difference of the operators from ones in the Schr\"{o}dinger equation consists in the fact that the wave functions applied by them are defined in phase space. 
For the Schr\"{o}dinger equation, the wave function is the state function defined in configuration space, whereas for the master equation of QMPS the wave function is the state function defined in phase space.
 
Thus, the multiplication by $i\hbar$  helps us to conceive the conception of operators as the tools for determining physical quantities in the quantum world in terms of the wave function.
It is emphasized that the dynamical quantities obtained with the help of operators and wave function are not the same as classical ones, and get quantal. 

Such an interpretation on quantum observables naturally leads to formulating time as an ordinary quantum observable.
If extending the phase space furthermore, we take the action as
\begin{equation}\label{eq:44-2-2}
S\left(\mathbi{q}, \mathbi{p}, H, t\right)=\int_0^{\mathbi{q}} \mathbi{p}~ \mathrm{d}\mathbi{q}' -\int_0^t H \mathrm{d}t'~,
\end{equation}
then we get
\begin{equation}\label{eq:44-2-3}
\frac{\partial S\left(\mathbi{q}, \mathbi{p}, H, t\right)}{\partial H} =-t.
\end{equation}
Therefore the time operator becomes 
\begin{equation}
\hat{t}=i\hbar \frac{\partial}{\partial H}~.
\end{equation}
Thus, the special status of time as an observable comes to be lost.

The introduction of these operators helps to clarify the relations of this formalism with the others.
Using the above operators, we can write Eq. \ref{eq:42_37} as 
\begin{equation}\label{eq:44_3}
\boxed{\hat{E}\psi=\frac{1}{2} \left\{\sum_{i=1}^{f} \left(\dot{q}_i \hat{p}_i+\dot{p}_i \hat{q}_i\right)\right\}\psi}~,
\end{equation}
in more compact form,
\begin{equation*}\label{eq:46}
\hat{E} \psi= \hat{H} \psi~,
\end{equation*}
where the Hamiltonian operator takes the following form: 
\begin{equation}\label{eq:47}
\hat{H}= \hat{T}+\hat{U}~.
\end{equation}
Although trivial and even meaningless in the mathematical aspect, in order both to put special emphasis on operator and to make it heuristic, we adopt Eq. \ref{eq:42_37} as the master equation of QMPS. 
Of course, Eq. \ref{eq:41} is identical to Eq. \ref{eq:42_37}.
The master equation of this formalism is distinguished from the Schr\"{o}dinger equation because the wave function is defined not in configuration space but in phase space. 
This formalism is expected to be useful to elucidate the relations to the other formulations of quantum mechanics. 

\subsection{\label{conc_form}Time-independent equation}

We shall deduce time-independent wave equation from Eq. \ref{eq:42_37} on the basis of the assumption on the form of wave function.

Recall the assumption about the form of wave function,
\begin{align}\label{eq:49}
\psi \left(\mathbi{q},\mathbi{p},t\right)&=\phi \left( \mathbi{q},\mathbi{p},t \right) \exp{\left(\frac{i}{\hbar}S\right)}\nonumber \\
& =\phi \left(\mathbi{q},\mathbi{p},t\right) \exp \left\{\frac{i}{\hbar} \left(\int{\mathbi{p} \mathrm{d} q}-\int H \mathrm{d} t\right)\right\} \nonumber\\
& =\varphi \left(\mathbi{q},\mathbi{p},t \right) \exp \left( -\frac{i}{\hbar}\int H \mathrm{d} t\right)~,
\end{align}
where $\phi \left( \mathbi{q},\mathbi{p},t \right)$ is a real-valued function.

In the time-independent case, since the Hamiltonian function explicitly does not involve time variable, it becomes the integration of motion.
Accordingly,
\begin{equation*}\label{eq:49_a}
H\left(\mathbi{q},\mathbi{p}\right)=E=const~.
\end{equation*}

Therefore the wave function is of the following form:
\begin{equation}\label{eq:49_b}
\psi \left(\mathbi{q},\mathbi{p},t\right)=\varphi \left(\mathbi{q},\mathbi{p},t \right) \exp \left( -\frac{i}{\hbar} Et\right).
\end{equation}
Inserting Eq. \ref{eq:49_b} into the time-dependent wave equation yields
\begin{eqnarray*}\label{eq:50}
i\hbar \left\{\varphi \left(\mathbi{q},\mathbi{p},t \right)\left(- \frac{i}{\hbar}E \right)+\frac{\partial}{\partial t}\varphi \left(\mathbi{q},\mathbi{p},t \right)\right\} exp \left(- \frac{i}{\hbar} Et \right) \nonumber \\
= - \frac{i\hbar}{2} \sum_{i=1}^{f}\left[ \frac{\partial H}{\partial p_i} \frac{\partial \varphi}{\partial q_i}-\frac{\partial H}{\partial q_i}\frac{\partial \varphi}{\partial p_i}  \right] \cdot exp \left( - \frac{i}{\hbar} Et \right)~,
\end{eqnarray*}
then we have
\begin{align}\label{eq:51}
&E \varphi \left(\mathbi{q},\mathbi{p},t \right)+i\hbar \frac{\partial}{\partial t}\varphi \left(\mathbi{q},\mathbi{p},t \right)\nonumber\\
&=-\frac{i\hbar}{2} \sum_{i=1}^{f}\left[ \frac{\partial H}{\partial p_i}\frac{\partial \varphi}{\partial q_i}-\frac{\partial H}{\partial q_i}\frac{\partial \varphi}{\partial p_i} \right]~.
\end{align}
From the above equation, we obtain the time-independent equation. 
The second term in the left side of Eq. \ref{eq:51} vanishes due to the time-independence of the amplitude function, $\phi \left(\mathbi{q},\mathbi{p},t \right)$. 

The explanation runs as follows.
If the probability density is time-independent, then the equation of movement for the probability density is represented as 
\begin{equation}\label{eq:add1}
\sum_{i=1}^{f}\left( \dot{q_i} \frac{\partial \rho}{\partial q_i}+\dot{p_i} \frac{\partial \rho}{\partial p_i} \right)=0~.
\end{equation}

The generalized solution to this equation takes the form of an arbitrary function with respect to the Hamiltonian function, i.e.
\begin{equation}
\label{eq:add2}
\rho=\rho\left\{ f \left[H\left(\mathbi{q}, \mathbi{p} \right) \right] \right\}~,
\end{equation}
where $ f $ is an arbitrary function.

Without loss of generality, it can be concluded that if the Hamiltonian function of a system under consideration is time-independent, then its probability density is given as a definite function dependent on the Hamiltonian function.
Therefore the time-independence of the Hamiltonian function indicates that of the probability density.
From this it follows that the amplitude function, $\phi \left(\mathbi{q},\mathbi{p},t \right)$ is time-independent and accordingly, the second term in the left side of Eq. \ref{eq:51} vanishes. 

Therefore we obtain the following time-independent equation:
\begin{equation}\label{eq:52}
-\frac{i\hbar}{2} \sum_{i=1}^{f}\left[ \frac{\partial H}{\partial p_i}\frac{\partial \varphi \left( \mathbi{q}, \mathbi{p} \right)} {\partial q_i}-\frac{\partial H}{\partial q_i} \frac{\partial \varphi \left( \mathbi{q},\mathbi{p} \right)}{\partial p_i}  \right] =E \varphi \left( \mathbi{q}, \mathbi{p} \right)~.
\end{equation}

On the other hand, it can be also shown that the time-independent wave equation is easily obtained by using the variable separation method. 
By separating the time variable, the wave function is written as 
\begin{equation}\label{eq:52_a}
\psi \left(\mathbi{q},\mathbi{p},t\right)=\varphi \left(\mathbi{q},\mathbi{p} \right) f \left(t\right).
\end{equation} 
Inserting this function into the time-dependent wave equation
\begin{equation}\label{eq:52_b}
i\hbar\frac{\partial \psi }{\partial t}=-\frac{i\hbar}{2}\sum_{i=1}^{f} \left[ \frac{\partial H}{\partial p_i} \frac{\partial \psi}{\partial q_i}-\frac{\partial H}{\partial q_i} \frac{\partial \psi}{\partial p_i} \right]~,
\end{equation}
we obtain
\begin{align}\label{eq:52_c}
&i\hbar \varphi\left(\mathbi{q},\mathbi{p}\right)\frac{\partial f\left(t\right)}{\partial t}\nonumber\\
&=-\frac{i\hbar}{2}f\left(t\right)\sum_{i=1}^{f} \left[ \frac{\partial H}{\partial p_i} \frac{\partial \varphi\left(\mathbi{q},\mathbi{p}\right)}{\partial q_i}-\frac{\partial H}{\partial q_i} \frac{\partial \varphi\left(\mathbi{q},\mathbi{p}\right)}{\partial p_i} \right]~.
\end{align}
Since for stationary states the Hamiltonian function is independent of time, dividing the both sides of the above equation by $\varphi \left(\mathbi{q},\mathbi{p} \right) f \left(t\right)$ leads to variable separation.
Thus, we have
\begin{align}\label{eq:52_d}
&i\hbar \frac{1}{f\left(t\right)}\frac{\partial f\left(t\right)}{\partial t}\nonumber\\
&=-\frac{i\hbar}{2}\frac{1}{\varphi\left(\mathbi{q},\mathbi{p}\right)}\sum_{i=1}^{f} \left[ \frac{\partial H}{\partial p_i} \frac{\partial \varphi\left(\mathbi{q},\mathbi{p}\right)}{\partial q_i}-\frac{\partial H}{\partial q_i} \frac{\partial \varphi\left(\mathbi{q},\mathbi{p}\right)}{\partial p_i} \right]~.
\end{align}
As a consequence, we obtain the wave equation with respect to time,
\begin{equation}\label{eq:52_e}
i\hbar\frac{\partial f\left(t\right)}{\partial t}=E f\left(t\right)	~.
\end{equation}
This time-dependent wave equation easily yields the solution 
\begin{equation}\label{eq:52_f}
f\left(t\right)=c \exp{\left(-\frac{i}{\hbar}Et\right)}~.
\end{equation}
From Eq. \ref{eq:52_d}, we get the time-independent equation
\begin{equation}\label{eq:52_g}
\boxed{-\frac{i\hbar}{2}\sum_{i=1}^{f} \left[ \frac{\partial H}{\partial p_i} \frac{\partial \varphi\left(\mathbi{q},\mathbi{p}\right)}{\partial q_i}-\frac{\partial H}{\partial q_i} \frac{\partial \varphi\left(\mathbi{q},\mathbi{p}\right)}{\partial p_i} \right]=E\varphi\left(\mathbi{q},\mathbi{p}\right)}~.
\end{equation}

Thus, the concrete forms of both time-dependent and time-independent equation have been reviewed.

\section{\label{rel_mas_eq}Relation between formalism of QMPS and the Schr\"{o}dinger equation, probabilistic and mechanical structure of formalism}
\subsection{\label{rel_mas_sch_eq}Relation between master equation of QMPS and the Schr\"{o}dinger equation}

Introducing the conception of operators into the equation of QMPS  enables us to easily interpret the relation between the master equation of QMPS and the Schr\"{o}dinger equation.
We start with the momentum operator.
For a given wave function, $\psi=\psi_o \exp{\dfrac{i}{\hbar} S}$, the momentum operator, $ \hat{\mathbi{p}}$ should satisfy the following operator equation:
\begin{eqnarray*}\label{eq:53}
\hat{\mathbi{p}}\psi=\mathbi{p} \psi + \frac{1}{\psi_o}\hat{\mathbi{p}} \psi_o \cdot \psi=\left( \mathbi{p} + \frac{1}{\psi_o}\hat{\mathbi{p}} \psi_o \right) \psi =
\tilde{\mathbi{p}} \psi~,
\end{eqnarray*}
where $\psi_o$ is a real-valued function and in general, $\tilde{\mathbi{p}}$, a complex-valued function.
In the above equation we take into consideration $ \hat{\mathbi{p}}\left(\exp{\dfrac{i}{\hbar} S}\right )=\mathbi{p}\cdot \exp{\dfrac{i}{\hbar} S}$.

Therefore we can imagine the following correspondence:
\begin{equation*}\label{eq:a53}
\mathbi{p}_R=Re \left(\frac{1}{\psi} \hat{\mathbi{p}} \psi \right)=\mathbi{p}~,
\end{equation*}

\begin{equation*}\label{eq:a54}
\mathbi{p}_I=Im \left(\frac{1}{\psi} \hat{\mathbi{p}} \psi \right)=\frac{-i}{\psi_o} \hat{\mathbi{p}} \psi_o~.
\end{equation*} 

Without loss of generality, we have $\tilde{\mathbi{p}}=\mathbi{p} + \dfrac{1}{\psi_o}\hat{\mathbi{p}} \psi_o =\mathbi{p}_R+i\mathbi{p}_I$.
 
The above operator equation shows how the operator corresponding to a given dynamical quantity should yield the corresponding dynamical quantity as a result of its application to the wave function. 
In general, dynamical quantities in quantum mechanics are represented not by an eigenvalue, but by a function. 
In the above expression, $\mathbi{p}$ generally is not an eigenvalue but a function pertaining to observables. 
Therefore the function, $\mathbi{p}$ can be referred to as the function on momentum. 
On the other hand, $ \mathbi{p}_I$ can be considered to be relevant to the wave-like characteristics.

Now, we consider the approximation of the master equation to the Schr\"{o}dinger equation.
Introducing the momentum instead of the phase velocity in Eq. \ref{eq:42_37} gives
\begin{align}\label{eq:55}
-\frac{i\hbar}{2} \sum_{i=1}^{f}\left[ \frac{p_i}{m_{(p_i)}} \frac{\partial \Phi\left( \mathbi{q},\mathbi{p},t \right)}{\partial q_i} -\frac{\partial H}{\partial q_i} \frac{\partial \Phi \left( \mathbi{q},\mathbi{p},t \right)}{\partial p_i} \right]\nonumber\\
=i \hbar \frac{\partial}{\partial t} \Phi \left( \mathbi{q},\mathbi{p},t \right)~. 
\end{align}
Here, $\Phi$ is the wave function and $m_{(p_i)}$ denotes the mass of a particle having the momentum component, $p_i$.

Corresponding $-i\hbar \dfrac{\partial}{\partial q_i}$ to the momentum operator, $\hat{p}_i$ to  leads to
\begin{align}\label{eq:56}
\frac{1}{2} \sum_{i=1}^{f}\left[ \frac{p_i}{m_{(p_i)}} \hat{p_i}+i \hbar \frac{\partial H}{\partial q_i} \frac{\partial}{\partial p_i} \right]\Phi \left( \mathbi{q},\mathbi{p},t \right)\nonumber\\
=i \hbar \frac{\partial}{\partial t} \Phi \left( \mathbi{q},\mathbi{p},t \right)~.
\end{align}

By assumption, we write $\Phi \left(\mathbi{q}, \mathbi{p}, t \right)=\Phi_o \left(\mathbi{q}, \mathbi{p}, t \right)\cdot \exp{\dfrac{i}{\hbar} S}$.
Here $\Phi_o \left(\mathbi{q}, \mathbi{p}, t \right)$ is a real-valued function.
Taking into account the commutation relation between momentum and momentum operator, and the operator equation,
\begin{eqnarray*}\label{eq:59}
\hat{p}_i \Phi \left(\mathbi{q}, \mathbi{p}, t \right)=p_i \Phi \left(\mathbi{q}, \mathbi{p}, t \right)+ \frac{1}{\Phi_o}\hat{p}_i \Phi_o \cdot \Phi~, 
\end{eqnarray*}
we get the following equation:
\begin{align}\label{eq:60_51}
\left[\frac{1}{2} \sum_{i=1}^{f}\frac{\hat{p}_i^2}{m_{(p_i)}}- \frac{1}{2} \sum_{i=1}^{f}\frac{1}{m_{(p_i)}}\hat{p}_i \left(\frac{1}{\Phi_o}\hat{p}_i \Phi_o \right)+\hat{U} \right] \Phi \left(\mathbi{q}, \mathbi{p}, t \right)\nonumber\\
=i \hbar \frac{\partial}{\partial t} \Phi \left(\mathbi{q}, \mathbi{p}, t \right)~.
\end{align}

For the sake of convenience, we introduce following notation:
\begin{equation}\label{eq:60_52} 
\hat{U}_o = -\frac{1}{2} \sum_i \frac{1}{m_{(p_i)}}\hat{p}_i \left(\frac{1}{\Phi_o}\hat{p}_i \Phi_o \right)+\hat{U} 
\end{equation}

To transform Eq. \ref{eq:60_51} into Schr\"{o}dinger equation, it is necessary to perform the following variables separation:
\begin{equation}\label{eq:61}
\Phi \left(\mathbi{q}, \mathbi{p}, t \right)=\psi \left(\mathbi{q}, t \right) \phi \left(\mathbi{p} \right)~.
\end{equation}

Inserting the above function into Eq. \ref{eq:60_51}, we get
\begin{eqnarray*}\label{eq:62}
\left( \sum_{i=1}^{f}\frac{\hat{p}_i^2}{2 m_{(p_i)}}+\hat{U}_o \right) \left[ \psi \left(\mathbi{q}, t \right) \phi \left(\mathbi{p} \right) \right]=i \hbar \frac{\partial}{\partial t} \left[ \psi \left(\mathbi{q}, t \right) \phi \left(\mathbi{p} \right) \right]~.
\end{eqnarray*}

Multiplying both sides of the above equation by $\phi^\ast \left( \mathbi{p} \right)$  and integrating it over $\mathbi{p}$, we obtain the following equation:
\begin{eqnarray}
\label{eq:63}
\left( \sum_{i=1}^{f}\frac{\hat{p}_i^2}{2 m_{(p_i)}}+ \int \phi^\ast \hat{U}_o \phi \mathrm{d} \mathbi{p} \right) \psi \left(\mathbi{q}, t \right) =i \hbar \frac{\partial}{\partial t} \psi \left(\mathbi{q}, t \right). 
\end{eqnarray}

In the above calculation, we made use of the following normalization condition:
\begin{equation*}
\label{eq:64}
\int{\phi^\ast \phi \, \mathrm{d} \mathbi{p}}=1~.
\end{equation*}

The approximation of the integral expression relevant to the potential operator, $\int \phi^\ast \hat{U}_o \phi \mathrm{d} \mathbi{p}$ to the potential function,
\begin{equation*}\label{eq:65}
\int \phi^\ast \hat{U}_o \phi \, \mathrm{d} \mathbi{p}=U 
\end{equation*}
yields the Schr\"{o}dinger equation in configuration space,
\begin{equation}\label{eq:66}
\left( \sum_{i=1}^{f}\frac{\hat{p}_i^2}{2 m_{(p_i)}}+U \right) \psi \left(\mathbi{q}, t \right)=i \hbar \frac{\partial}{\partial t} \psi \left(\mathbi{q}, t \right)~. 
\end{equation}

Thus, we explain that the master equation of QMPS approximates to the Schr\"{o}dinger equation as a special case.

\subsection{\label{subsec:App to CM} Probabilistic and mechanical structure of formalism}

In order to demonstrate the mechanical structure of QMPS, it is enough to show that the master equation of QMPS contains such an energy relation of particles as one in classical mechanics.

Starting with the form of wave function, we review the structure of the master equation of QMPS. 

Substituting the wave function, 
\begin{equation}\label{eq:67}
\psi \left( \mathbi{q}, \mathbi{p},t \right)=\psi_0 \left( \mathbi{q}, \mathbi{p},t \right) \cdot \exp{\left(\dfrac{iS}{\hbar} \right)} 
\end{equation}
into the master equation of QMPS and separating the equation into real and imaginary part, we get 
\begin{equation}
\sum_{i=1}^{f}\left( \dot{q}_i \frac{\partial \psi_0^2}{\partial q_i} + \dot{p}_i \frac{\partial \psi_0^2}{\partial p_i} \right)= \frac{\partial \psi_0^2}{\partial t}~,
\end{equation}
i.e.
\begin{equation}\label{eq:68}
\sum_{i=1}^{f}\left( \dot{q}_i \frac{\partial \rho}{\partial q_i} + \dot{p}_i \frac{\partial \rho}{\partial p_i} \right)=
\frac{\partial \rho}{\partial t}~,
\end{equation}
and
\begin{equation}\label{eq:69}
\sum_{i=1}^{f}{\left( \dot{q}_i \frac{\partial S}{\partial q_i}+\dot{p}_i \frac{\partial S}{\partial p_i} \right) }+\frac{\partial S}{\partial t}= 0 ~,
\end{equation}
i.e.
\begin{equation}\label{eq:70}
\sum_{i=1}^{f}{\left( \frac{p^2_i}{2m_{(p_i)}}-\frac{1}{2}\,q_i\,\frac{\partial H}{\partial q_i}  \right) }= H ~.
\end{equation}
Here $\psi_0$ is a real-valued function.

Eq. \ref{eq:68} is the probability continuity equation, whereas Eq.  \ref{eq:70} represents the energy relation of particles. 
Eq. \ref{eq:69} pertaining to the phase relation implies that there is no change in phase of probability wave with respect to a system moving at the phase velocity. 
The wave equation thus involves the duality, i.e. the particle-like and wave-like property. 
Exactly, the wave equation contains the relations of not only energy of particles but also probability wave. 

The second term of Eq. \ref{eq:70},

\begin{equation}\label{eq:71}
U_{quant}=-\frac{1}{2} \sum_{i=1}^{f}\,q_i\,\frac{\partial H}{\partial q_i}
\end{equation}

means that the potential of a quantal entity is different from the classical potential. 
It is a matter of course that the quantal potential should be represented as other than the classical potential. 
In fact, if the quantal potential were the same as classical one, one could not find any other behaviors of microparticles than classical ones.
The quantum potential is represented by the product of the force acting on particles and the position vector determined in terms of the action. 

It is possible to explain the quantum potential in terms of QMPS as in Bohmian mechanics \cite{Bowman}.
For the master equation of QMPS, the potential operator becomes a composite operator composed of differentiation and multiplicand. 
This composite operator should be considered as the potential operator that reflects both classical and quantal causality.
Therefore the potential displaying the pure quantal causality can be denoted by

\begin{align}\label{eq:83}
U_q=\sum_{i=1}^{f}\mathrm{Re} \left(\frac{i \hbar}{2 \psi} \frac{\partial H}{\partial q_i} \frac{\partial}{\partial p_i} \psi \right)-U\nonumber\\
= \sum_{i=1}^{f} \mathrm{Re} \left(\frac{i \hbar}{2} \frac{\partial H}{\partial q_i} \frac{\partial}{\partial p_i} \ln \psi \right) -U  ~.
\end{align}
Hence, the quantal force can be determined by 
\begin{equation*}\label{eq:84}
\mathbi{F}_q=- \nabla U_q ~.
\end{equation*}

It is conceivable that this quantal force causes some quantal fluctuation in the classical paths of microparticles. 

For example, We consider the problem for a free particle. 
A free particle is considered to move along a classical trajectory even though it is a quantum-mechanical particle. 
According to the standard interpretation of quantum mechanics, the wave function for a free particle is interpreted to be distributed over whole space. 
This interpretation is not compatible with physical reality. 
The case shows that the wave function cannot provide the complete information of a quantal system. 
The view of standard theory of quantum mechanics is that the wave function provides complete information of a quantum system. 
In contrast, the causal theory of quantum mechanics claims that the wave function cannot provide complete information, but amounts to the tools for obtaining complete dynamical information on a quantum system. 
Obviously, the quantum force affecting a free particle in terms of  QMPS is zero.  
Therefore a free particle, though it is a quantum object, behaves like a classical particle without undergoing any quantal fluctuations. 

Furthermore, we shall consider the case when the potential of a given field is homogeneous in some directions.
For example, let the potential be $U=U\left(x,\, t\right)$.
Consequently, the potential is independent of $y$ and $z$.
Then the wave equation is written as 
\begin{eqnarray}\label{eq:83a}
\left( \sum_{i=1}^{3}\frac{p_i\hat{p}_i}{2 m}-\frac{1}{2}\frac{\partial}{\partial q_i}U\left(x,t\right) \hat{q}_i \right) \psi \left(\mathbi{q}, \mathbi{p}, t \right)\nonumber\\
=i \hbar \frac{\partial}{\partial t} \psi \left(\mathbi{q}, \mathbi{p}, t \right)~. 
\end{eqnarray}
In this case, the above equation is separated with respect to variables as
\begin{subequations}
\begin{gather}
\label{eq:83b}
\frac{p_x\hat{p}_x}{2 m}\psi_x \left(x, p_x, t \right)-\frac{1}{2}\frac{\partial}{\partial x}U\left(x, t\right) \hat{x} \psi_x \left(x, p_x, t \right)\nonumber\\
=i \hbar \frac{\partial}{\partial t} \psi_x \left(x, p_x, t \right)~,\\
\label{eq:83c}
\frac{p_y\hat{p}_y}{2 m} \psi_y \left(y, p_y, t \right)=i \hbar \frac{\partial}{\partial t} \psi_y \left(y, p_y, t \right)~,\\
\label{eq:83d}
\frac{p_z\hat{p}_z}{2 m}\psi_z \left(z, p_z, t \right)=i \hbar \frac{\partial}{\partial t} \psi_z \left(z, p_z, t \right)~,
\end{gather}
\end{subequations} 
where 
\begin{eqnarray}\label{eq:83e}
&\psi \left(x, y, z, p_x, p_y, p_z, t \right)\nonumber\\
&=\psi_x \left(x, p_x, t \right)\psi_y \left(y, p_y, t \right)\psi_z \left(z, p_z, t \right).
\end{eqnarray}  
Equations \ref{eq:83c} and \ref{eq:83d} are identical to the wave equation for a free particle.
In this case, the quantal forces via Eq.\ref{eq:83} equal zero.
Therefore the particle does not undergo any fluctuation in both $y$ and $z$ directions, and as a result, there is no spread of its trajectory.

\section{\label{sec:application}Application to some simple cases: verification of validity}
\subsection{\label{free_particle}Free particle problem}

We consider the solutions to wave equations for simple cases with the help of the master equation of QMPS.
The application examples will be an important test for verifying validity of the formalism.
For the sake of simplicity, we shall consider the problem for a free particle in one-dimensional case. 

In this case, the master equation of QMPS represented with respect to phase velocity is 
\begin{equation}\label{eq:87}
\frac{\partial \psi}{\partial t}+v_x \frac{\partial \psi}{\partial x}=0 ~.
\end{equation}

The solution to this equation is
\begin{equation}\label{eq:88}
\psi=\Phi \left( x-v_x t \right) ~,
\end{equation}
where $\Phi$ designates an arbitrary function.

If a certain number k is introduced, then it turns into
\begin{equation}\label{eq:89}
\psi=\Phi \left( kx-kv_x t \right)~.
\end{equation}

Setting $kv_x=\omega$, we have 
\begin{equation*}\label{eq:90}
k= \frac{\omega}{v_x} ~.
\end{equation*}

Obviously, it indicates the wave vector.
Introducing this relation yields the following wave function:
\begin{equation*}\label{eq:91}
\psi=\Phi \left( kx-\omega t \right)~.
\end{equation*}

In view of de Broglie's relation, we get 
\begin{equation*}\label{eq:92}
\psi=\Phi \left( px-E t \right)~. 
\end{equation*}

Recall that we already assumed
\begin{equation*}\label{eq:93}
\psi_1=\psi_0 \left( \mathbi{q},\mathbi{p},t \right) \exp{\left(\frac{iS}{\hbar}\right)}
\end{equation*}
for the form of wave function.

From this assumption, the concrete form of the wave function for a free particle can be represented as
\begin{equation}\label{eq:94}
\psi=A \exp \left\{\frac{i}{\hbar} \left( px-Et \right)\right\}~.
\end{equation}

The above expression shows that the new equation exactly describes free particles. 
Inserting the wave function into Eq. \ref{eq:87} proves the energy relation for a free particle. 
Thus, it can be seen that the wave function for a free particle is represented by a harmonic function. 

\subsection{\label{harmo_oscil}Harmonic oscillator problem}

The harmonic oscillator plays a prominent role in physics, in particular, in quantum mechanics.
Since it can be solved exactly, it may provide the clue to validity of an alternative formalism.
We treat the problem of harmonic oscillator to validate this formalism.
We start from the equation for one-dimensional harmonic oscillator,
\begin{equation*}\label{eq:95}
- \frac{i \hbar}{2} \left[ \frac{\partial \psi}{\partial x} \frac{\partial H}{\partial p_x}-\frac{\partial \psi}{\partial p_x} \frac{\partial H}{\partial x} \right]=E \psi~,
\end{equation*}
\begin{equation*}\label{eq:96}
\frac{1}{2} \left[ \frac{\partial \psi}{\partial x} \frac{\partial H}{\partial p_x}-\frac{\partial \psi}{\partial p_x} \frac{\partial H}{\partial x} \right]=\frac{iE}{\hbar}\psi~.
\end{equation*}
Inserting the Hamiltonian function of harmonic oscillator,
\begin{equation*}\label{eq:97}
H= \frac{p_x^2}{2 m}+\frac{m \omega^2}{2}x^2
\end{equation*}
into the time-independent wave equation gives the following result:
\begin{equation*}\label{eq:98}
\frac{\partial H}{\partial p_x}=\frac{p_x}{m}, \: \: \frac{\partial H}{\partial x}=m \omega^2 x~,
\end{equation*}
\begin{equation}\label{eq:99_90}
\frac{p_x}{m}\frac{\partial \psi}{\partial x}-m \omega^2 x \frac{\partial \psi}{\partial p_x}=\frac{2iE}{\hbar} \psi~.
\end{equation}
Dividing both sides of Eq. \ref{eq:99_90} by $\omega$, we have the following equation:
\begin{equation*}\label{eq:100}
\frac{\partial \psi}{\partial x} \frac{p_x}{m \omega}-m \omega x \frac{\partial \psi}{\partial p_x}=\frac{2iE}{\hbar \omega} \psi~.
\end{equation*}
Instead of $m \omega x$ introducing $\xi $ gives the following equation:
\begin{equation*}\label{eq:101}
p_x \frac{\partial \psi}{\partial \xi}-\xi \frac{\partial \psi}{\partial p_x}=\frac{2iE}{\hbar \omega} \psi~.
\end{equation*}

From the characteristic equation for first-order partial differential equations, we get the following equality:
\begin{equation}\label{eq:102}
\dfrac{\mathrm{d}\xi}{p_x}=-\dfrac{\mathrm{d}p_x}{\xi}=\dfrac{\mathrm{d}\psi}{\dfrac{2iE}{\hbar \omega}\psi}~.
\end{equation}

For the sake of simplicity, we introduce the notation,
\begin{equation*}\label{eq:103}
\frac{2E}{\hbar \omega}=\bar{n}~.
\end{equation*}

It is possible to prove that $\bar{n}$, initially assumed as an arbitrary number, should by all means be an integer.
From the successive two expressions of ~Eq. \ref{eq:102}, we have
\begin{equation*}\label{eq:104}
\frac{\mathrm{d}\xi}{p_x}=-\frac{\mathrm{d}p_x}{\xi}~,
\end{equation*}
\begin{equation*}\label{eq:105}
\mathrm{d} \left( \frac{\xi^2+p_x^2}{2} \right)=0~,
\end{equation*}
\begin{equation*}\label{eq:106}
\xi^2+p_x^2=c_1~,
\end{equation*}
\begin{equation*}\label{eq:107}
p_x=\pm \sqrt{c_1-\xi^2}~.
\end{equation*}

On the other hand, from the first expression and the last one of ~Eq. \ref{eq:102}, we get
\begin{equation*}\label{eq:108}
\ln \psi = \pm i \bar{n} \int \frac{1}{\sqrt{c_1}} \frac{\mathrm{d} \xi}{\sqrt{1-\left( \frac{\xi}{\sqrt{c_1}}\right)^2}}~.
\end{equation*}
Hence, we obtain
\begin{equation*}\label{eq:109}
\ln \psi=\pm i \bar{n} \arcsin \frac{\xi}{\sqrt{c_1}} +c_2~.
\end{equation*}
Since the generalized solution to a first-order partial differential equation is obtained as an arbitrary function of the constants given through the solutions to characteristic equations, a generalized solution to the equation is 
\begin{align}\label{eq:111}
\psi = c \exp \left( \pm i \bar{n} \arcsin \frac{m \omega x}{\sqrt{m^2 \omega^2 x^2 + p_x^2}} \right) \cdot \nonumber\\
\exp \left\{ \Phi \left( \frac{m^2 \omega^2 x^2 + p_x^2 }{2} \right) \right\}~,
\end{align}
where $\Phi$ is an arbitrary function. 
Therefore a possible solution satisfying the finiteness of wave function can be chosen as

\begin{eqnarray}\label{eq:112}
\psi = c \exp \left( \pm i \bar{n} \arcsin \dfrac{m \omega x}{\sqrt{2 m}\sqrt{\dfrac{m \omega^2 x^2}{2} + \dfrac{p_x^2}{2 m}}} \right)\cdot \nonumber \\
\exp \left[ - \dfrac{1}{2 \beta}\left( m \omega^2 x^2 + \dfrac{p_x^2}{m} \right) \right]~,
\end{eqnarray}
where $\beta$ is a constant.

It is possible to derive the quantization results by imposing
appropriate boundary conditions. 
For standard theory of quantum mechanics, the concept of the conversion point of harmonic oscillators is meaningless. 
That is because overall-space distribution of quantum entity is assumed. 
In the case of harmonic oscillator, the problem arises from assuming only infinite amplitude. 
For the standard theory of quantum mechanics the quantization results are obtained from the finiteness condition for a wave function. Therefore we cannot but admit the wave function to extend beyond the classical limit of values of coordinates \cite{Whitaker}. 
We conceive the conversion point even for quantum-mechanical harmonic oscillators. 
This point determines the limitation of distribution. 
In this case, $p_x$ should vanish at the conversion point, $a$. 

Hence we get
\begin{equation*}\label{eq:113}
\psi \vert_{x=a}=c \exp{\left(\pm i \bar{n} \arcsin{1} \right)} \cdot \exp{\left(- \frac{m \omega^2 a^2}{2 \beta} \right)}~.
\end{equation*}

There are two possible cases.
For the first case, we have
\begin{equation*}\label{eq:114}
\psi \vert_{x=a}=A \cos{\left(\bar{n} \frac{\pi}{2} \right)} \cdot \exp{\left(- \frac{m \omega^2 a^2}{2 \beta} \right)}~.
\end{equation*}
Then the boundary condition should be
\begin{equation*}\label{eq:116}
\psi \vert_{x=a}=0~.
\end{equation*}
From this condition, it follows that
\begin{equation*}\label{eq:117}
\bar{n} \frac{\pi}{2}=\left( n+1/2 \right) \pi ~.
\end{equation*}
Therefore we obtain the quantized energy of a harmonic oscillator,
\begin{equation*}\label{eq:118}
E_n=\left( n+ \frac{1}{2} \right) \hbar \omega~.
\end{equation*}

For the second case, we have
\begin{equation*}\label{eq:115}
\psi \vert_{x=a}=B \sin{\left(\bar{n} \frac{\pi}{2} \right)} \cdot \exp{\left(- \frac{m \omega^2 a^2}{2 \beta} \right)}~.
\end{equation*}
Then we have
\begin{equation*}\label{eq:119}
\bar{n} \frac{\pi}{2}=n \pi~.
\end{equation*}
From this, the quantized energy is denoted by
\begin{equation*}\label{eq:120}
E_n=n \hbar \omega~,
\end{equation*}
where $n$ is a positive integer.

As a result, the energy of a harmonic oscillator is 
\begin{equation}\label{eq:121}
E_n=\left( n+ \frac{1}{2} \right) \hbar \omega
\end{equation}
or
\begin{equation}\label{eq:122}
E_n=n \hbar \omega~.
\end{equation}

It is noteworthy that such quantization results do not depend upon the conversion point, or the amplitude of a harmonic oscillator. 
The solution to the equation of QMPS for a harmonic oscillator shows the same result as the Schr\"{o}dinger equation produces, since the eigenvalues of energy are quantized by $\hbar \omega$. 
As a result, we can verify that the master equation of QMPS gives reasonable results for harmonic oscillator.

\subsection{\label{subsec:two slits} Explanation of two-slit interference}
The two-slit experiment is at the core of the mysteries surrounding quantum mechanics.
Our view on this phenomenon is that since the wave field surrounding a particle is non-local, it passes through two slits and then the field disturbed by the two slits affects the movement of the particle afterward. 
All told, the particle passes through either of the two slits, while the wave field passes through both of the slits.
Naturally, particles are self-interferential because particle and field are inseparably unified. 
In essence, the diffraction of a particle through two slits is identified with the scattering problem.
Since microparticle is non-local, two slits can be regarded as a single unified object scattering an incoming particle.
In this case, the wave equation for explaining two-slit experiment writes as follows.
\begin{equation}\label{eq:123}
\frac{\partial \psi }{\partial t}=-\frac{1}{2}\sum_{i=1}^{3}\left[ \frac{\partial H}{\partial p_i} \frac{\partial \psi}{\partial q_i}-\frac{\partial H}{\partial q_i} \frac{\partial \psi}{\partial p_i} \right]~.
\end{equation}
Inserting the Hamiltonian into the above equation, we have
\begin{equation}\label{eq:124}
\frac{\partial \psi }{\partial t}=-\frac{1}{2}\sum_{i=1}^{3}\left[ \frac{p_i}{m} \frac{\partial \psi}{\partial q_i}-\frac{\partial U}{\partial q_i} \frac{\partial \psi}{\partial p_i} \right]~,
\end{equation}
where $U$ is the potential organized by two slits.
To unravel the two-slit diffraction in a fundamental way, it is necessary to find the sophisticated technique for determining $U$.

\begin{figure}[!th]
\centering
\includegraphics[clip=true, scale=0.45]{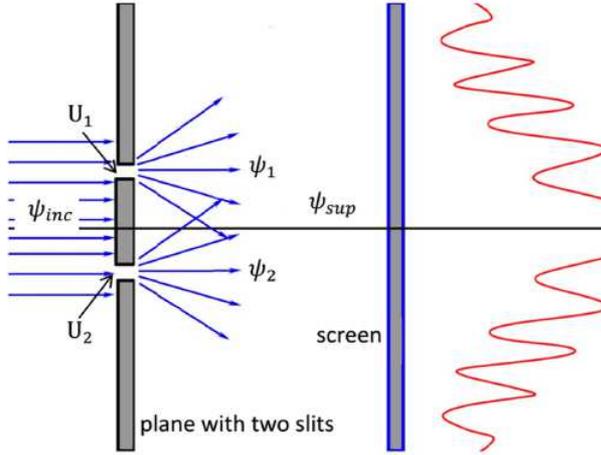}
\caption{\label{fig1}Two-slit experiment: $\psi_{inc}$ denotes the wave function of an incoming particle. $\psi_1$ is the wave function of the particle scattered by the potential $U_1$ and $\psi_2$, that by the potential $U_2$. The superposition of $\psi_1$ and $\psi_2$ furnishes the wave function of the scattered particle, $\psi_{sub}$.}
\end{figure}

Nevertheless it is possible to treat $U$ in a simple way.
It is justifiable to consider that $U$ approximates to $U_1$ or $U_2$  associated only with either of the two slits in the vicinity of it, respectively.
Therefore, solving eq. \ref{eq:124} we obtain two solutions, $\psi_1$ and $\psi_2$ for $U_1$ and $U_2$, respectively.
Since these are the solutions of eq. \ref{eq:124}, we can apply the principle of superposition.
Thus the scattered wave can be represented as the linear combination of $\psi_1$ and $\psi_2$.

Namely,
\begin{equation}\label{eq:125}
\psi_{sup}\left(\mathbi{q}, \mathbi{p}, t\right)=\psi_1\left(\mathbi{q}, \mathbi{p}, t\right)+\psi_2\left(\mathbi{q}, \mathbi{p}, t\right).
\end{equation}
The probability density is represented as
\begin{align}\label{eq:126}
&\rho\left(\mathbi{q}, \mathbi{p}, t\right)=\left|\psi_{sup}\left(\mathbi{q}, \mathbi{p}, t\right)\right|^2\nonumber\\
&=\left| \psi_1\left(\mathbi{q}, \mathbi{p}, t\right)+\psi_2\left(\mathbi{q}, \mathbi{p}, t\right) \right|^2\nonumber\\
&=\left| \psi_1\left(\mathbi{q}, \mathbi{p}, t\right)\right|^2+\left|\psi_2\left(\mathbi{q}, \mathbi{p}, t\right) \right|^2\nonumber\\
&~~~+\psi_1^*\left(\mathbi{q}, \mathbi{p}, t\right)\psi_2\left(\mathbi{q}, \mathbi{p}, t\right)+\psi_2^*\left(\mathbi{q}, \mathbi{p}, t\right)\psi_1\left(\mathbi{q}, \mathbi{p}, t\right)\nonumber\\
&=\left| \psi_1\left(\mathbi{q}, \mathbi{p}, t\right)\right|^2+\left|\psi_2\left(\mathbi{q}, \mathbi{p}, t\right) \right|^2\nonumber\\
&~~~+2\mathrm{Re}\left[\psi_1^*\left(\mathbi{q}, \mathbi{p}, t\right)\psi_2\left(\mathbi{q}, \mathbi{p}, t\right)\right].
\end{align}
The term, $2\mathrm{Re}\left[\psi_1^*\left(\mathbi{q}, \mathbi{p}, t\right)\psi_2\left(\mathbi{q}, \mathbi{p}, t\right)\right]$ represents the interference via the two slits.
The spatial distribution is determined by integration of $\rho\left(\mathbi{q}, \mathbi{p}, t\right)$ with respect to momentum $\mathbi{p}$.

Namely,
\begin{equation}\label{eq:126}
\rho_q\left(\mathbi{q}, t\right)=\int_{\mathbi{p}}\rho \left(\mathbi{q}, \mathbi{p}, t\right)\mathrm{d}\mathbi{p}.
\end{equation}
Here, we merely showed essentials for dealing with the two-slit interference.
One can obtain particular results, presupposing the initial wave function resulting from the effect of diffracting single slit on some incoming wave function\cite{Sbitnev,Sanz2}.
The superposition of the two waves coming from each slit gives rise to an interference pattern, which is modulated by the diffraction pattern associated with these slits.

\subsection{\label{subsec:Uncertain} Explanation of uncertainty relation}

The perspective of this formalism leads to understanding why the uncertainty relation exists. 
The consideration of the density of phase points forming a statistical ensemble along a phase trajectory indicates that phase volume occupied by it remains unchanged. 
For this reason, it turns out that the volume in configuration space occupied by a statistical ensemble of microparticles and that in momentum space are inversely proportional to each other. 
As a result, it follows that the higher the accuracy of position measurement is, the lower that of momentum measurement. 
This argument can be regarded as an account of the uncertainty relation which issues from the  point of view of statistical mechanics. 

Meanwhile, it is possible to consider the uncertainty relation from the commutation relation.
The momentum operator assumed in this formalism, which applies to the wave function in phase space, is in exact accord with that in the standard theory of quantum mechanics. 
Therefore it is concluded that for the two cases the commutation relation between position and momentum operator is identical. 
Therefore the uncertainty relation can be similarly proved for the two case. 

Now, we prove the uncertainty relation with regard to the wave function in phase space. 

With the help of the commutation relation between position and momentum operator,
\begin{equation*}\label{eq:128}
\left(x \hat{p}-\hat{p}x \right) =-i \hbar \left(x \frac{\partial }{\partial x}- \frac{\partial}{\partial x} x \right) =i \hbar ~,
\end{equation*}
we calculate 
\begin{align}\label{eq:129}
&\int_{-\infty}^\infty d p \int_{-\infty}^\infty d x \left| \alpha x \psi + \frac{\partial \psi}{\partial x} \right|^2  \nonumber \\
&=\int_{-\infty}^\infty d p \int_{-\infty}^\infty d x \left( \alpha x \psi^\ast + \frac{\partial \psi^\ast}{\partial x}\right) \left( \alpha x \psi + \frac{\partial \psi}{\partial x} \right) \nonumber \\
&= A \alpha^2 + B \alpha + C \geq 0~.
\end{align}
Hence we find
\begin{equation*}\label{eq:130}
B^2-4AC<0 \rightarrow \frac{B^2}{4}\leq AC~.
\end{equation*}
Then the calculated $A$, $B$ and $C$ are as follows:
\begin{eqnarray}\label{eq:131}
A=\int_{-\infty}^\infty dp \int_{-\infty}^\infty dx \alpha^2 x^2 \left| \psi \right|^2~,
\end{eqnarray}
\begin{align}\label{eq:133}
B&=\int_{-\infty}^\infty dp \int_{-\infty}^\infty dx \left(\psi x \frac{\partial}{\partial x} \psi^\ast 
+\alpha \psi^\ast x \frac{\partial}{\partial x} \psi \right)\nonumber\\
&=\int_{-\infty}^\infty dp \int_{-\infty}^\infty dx \psi^\ast \left( x \frac{\partial}{\partial x}-\frac{\partial}{\partial x} x \right) \psi\nonumber\\
&=\frac{-1}{i \hbar}\int_{-\infty}^\infty dp \int_{-\infty}^\infty dx \psi^\ast \left(x \hat{p}-\hat{p}x \right) \psi=-1~.
\end{align}
\begin{align}\label{eq:132}
C&=\int_{-\infty}^\infty dp \int_{-\infty}^\infty dx \frac{\partial \psi}{\partial x} \frac{\partial \psi^\ast}{\partial x}\nonumber\\
&=-\int_{-\infty}^\infty dp \int_{-\infty}^\infty dx \psi^\ast \frac{\partial^2 \psi}{\partial x^2}=\frac{\left<p^2\right>}{\hbar^2}~.
\end{align}
Hence it follows that the following relation:
\begin{equation}\label{eq:134}
\frac{1}{4}\leq \frac{\left< x^2 \right> \left< p^2 \right>}{\hbar^2} \Rightarrow \sqrt{\left< x^2 \right>} \cdot \sqrt{\left< p^2 \right>} \geq \frac{\hbar}{2}
\end{equation}
should hold. 

Thus, the uncertainty relation has been generally proved by means of the momentum operator in the present formalism.

\section{\label{sec:Extension}Relativistic wave equation: generalization and advance}

For this formalism, the phase velocity is considered as having real physical meaning.

Since the phase velocity plays an important role in the master equation of this formalism, it should be exactly explained whether or not it has physical meaning.
It is currently accepted that the phase velocity has not physical meaning for no other reason than for the relativistic case the phase velocity determined by definition exceeds the speed of light.
Meanwhile, it is inconsistent with our common sense in physics to accept the very fact that although the phase velocity is an important concept which characterizes the probability wave, it has no physical meaning.     
However such an inconsistency can be successfully resolved by proving that for the relativistic case the phase velocity cannot exceed the speed of light, provided that the rest energy of particle is considered as the origin of energy.
In relativistic mechanics, we define as the kinetic energy the part of the particle's energy that turns into zero as its velocity vanishes.   

Thus, we take the kinetic energy of a particle for
\begin{equation}\label{eq:135}
 K=\dfrac{m_0 c^2}{\sqrt{1-\dfrac{v^2}{c^2}}}-m_0 c^2~.
\end{equation}
Then the phase velocity is determined by
\begin{align}\label{eq:136}
 v_{phase}&=\dfrac{K}{p}=\dfrac{\dfrac{m_0 c^2}{\sqrt{1-\dfrac{v^2}{c^2}}}-m_0 c^2}{p}\nonumber\\
&=\dfrac{\dfrac{m_0 c^2-m_0 c^2 \sqrt{1-\dfrac{v^2}{c^2}}}{\sqrt{1-\dfrac{v^2}{c^2}}}}{\dfrac{m_0 v}{\sqrt{1-\dfrac{v^2}{c^2}}}}= \dfrac{c^2 \left(1-\sqrt{1-\dfrac{v^2}{c^2}} \right)}{v}~.
\end{align}

From this, it follows that as $ v \ll c $, $ v_{phase} \approx \frac{1}{2} v_{group} $, while as $ v \to c $, $ v_{phase}\to c $.
Meanwhile, as $ v \to 0 $, $ v_{phase}\to 0 $ according to L' Hospital rule.
Therefore it is no wonder to consider the phase velocity as a physically meaningful quantity. 

Now, we can go over to the problem of obtaining a relativistic wave equation.
To obtain the relativistic wave equation, we should replace $\dot{q}_i$ in Eq. \ref{eq:39} with the relativistic relation. 
Following the preceding argument about the phase velocity, we adopt Eq. \ref{eq:136} as the relativistic phase velocity.

Then for a particle, we have
\begin{equation}\label{eq:136a}
v_{phase}=\frac{K}{p}=\frac{\sqrt{c^2p^2-m_0^2c^4}-m_0c^2}{p}~.
\end{equation}
Expression \ref{eq:136a} can be represented in the vectorial form as 
\begin{equation*}\label{eq:136b}
\mathbi{v}_{phase}=\frac{\sqrt{c^2p^2-m_0^2c^4}-mc_0^2}{p^2}\mathbi{p}~.
\end{equation*}
Accordingly, the $i$th component of the phase velocity is written as 
\begin{equation}\label{eq:136c}
{\left(\mathbi{v}_{phase}\right)}_i=\frac{\sqrt{c^2p^2-m_0^2c^4}-m_0c^2}{p^2}~\left(p_i\right)_{curv}~,
\end{equation}
where $\left(p_i\right)_{curv}$ is the momentum component in a curvilinear coordinate system.
Note the relation between a generalized velocity $\dot{q}_i$ and the corresponding momentum component $\left(p_i\right)_{curv}$ in a curvilinear coordinate system,
\begin{equation}\label{eq:136d}
\left(p_i\right)_{curv}=m~g_i~\dot{q}_i=m~g_i~\frac{\partial H}{\partial p_i}~,
\end{equation}
and the relation between the generalized momentum component $p_i$ and momentum component ${\left(p_i\right)}_{curv}$ in a curvilinear coordinate system,
\begin{equation}\label{eq:136d}
p_i=g_i\cdot {\left(p_i\right)}_{curv}=m~g^2_i~\dot{q}_i~,
\end{equation}
where  $g_i$ is Lame's coefficient.

Accordingly, by use of the generalized momentum, we can write a component of generalized phase velocity as
\begin{align}\label{eq:136e}
&\left(\mathbi{v}_{phase}\right)_i^{(gen)}=\frac{\left(\mathbi{v}_{phase}\right)_i}{g_i}\nonumber\\
&=\frac{\sqrt{c^2p_{(p_i)}^2-m_{0(p_i)}^2c^4}-m_{0(p_i)}c^2}{p_{(p_i)}^2}~ \frac{(p_i)_{curv}}{g_i}~\nonumber\\
&=\frac{\sqrt{c^2p_{(p_i)}^2-m_{0(p_i)}^2c^4}-m_{0(p_i)}c^2}{p_{(p_i)}^2}~ \frac{p_i}{g_i^2}~,
\end{align}
where the term 'generalized' is abbreviated as $gen$, $p_{(p_i)}$  is the magnitude of the momentum whose one component corresponds  to $p_i$, and $m_{0(p_i)}$ is the rest mass of the particle corresponding to $p_i$.

By substituting the expression \ref{eq:136e} into Eq. \ref{eq:39}, we obtain
\begin{align}\label{eq:136f}
\frac{\partial \psi }{\partial t}=-\sum_{i=1}^{f} \left[ \frac{\sqrt{c^2p_{(p_i)}^2-m_{0(p_i)}^2c^4}-m_{0(p_i)}c^2}{p_{(p_i)}^2}~\frac{ p_i }{g_i^2}\frac{\partial \psi}{\partial q_i}\right.\nonumber\\
\left.-\frac{1}{2}\frac{\partial H}{\partial q_i}\frac{\partial \psi}{\partial p_i} \right]~.
\end{align}

Since the virial theorem is valid also for the relativistic case, the potential operator remains unchanged.
To represent the wave equation in terms of operators, we multiply the both sides of Eq. \ref{eq:136f} by $i\hbar$.
Then the wave equation is written as
\begin{align}\label{eq:136i}
i\hbar\frac{\partial \psi }{\partial t}=-i\hbar\sum_{i=1}^{f} \left[ \frac{\sqrt{c^2p_{(p_i)}^2-m_{0(p_i)}^2c^4}-m_{0(p_i)}c^2}{p_{(p_i)}^2}~\frac{p_i}{g_i^2}\frac{\partial \psi}{\partial q_i}\right.\nonumber\\
\left.-\frac{1}{2}\frac{\partial H}{\partial q_i}\frac{\partial \psi}{\partial p_i} \right]~,
\end{align}
\begin{align}\label{eq:136j}
\hat{E}\psi=\sum_{i=1}^{f} \left[ \frac{\sqrt{c^2p_{(p_i)}^2-m_{0(p_i)}^2c^4}-m_{0(p_i)}c^2}{p_{(p_i)}^2} \frac{p_i}{g_i^2}\hat{p_i} \psi\right.\nonumber\\
\left.-\frac{1}{2}\frac{\partial H}{\partial q_i} \hat{q_i}\psi \right]~.
\end{align}
Since for the relativistic case the negative energy is possible, the wave equation is finally represented as
\begin{align}\label{eq:136k}
\hat{E}\psi=\sum_{i=1}^{f} \left[\pm \frac{\sqrt{c^2p_{(p_i)}^2-m_{0(p_i)}^2c^4}-m_{0(p_i)}c^2}{p_{(p_i)}^2}~\frac{p_i}{g_i^2} \hat{p_i} \psi\right.\nonumber\\
\left.-\frac{1}{2}\frac{\partial H}{\partial q_i} \hat{q_i}\psi \right]~.
\end{align}

It is straightforward that for $c\gg v$, the relativistic wave equation \ref{eq:136f} turns into the non-relativistic wave equation \ref{eq:41}.
Therefore the present relativistic equation becomes the generalized wave equation of quantum mechanics in phase space. It implies an advance in combining the quantum theory with the relativity theory.

\section{\label{sec:IntAssum}Interpretation on assumption III}

It is possible to explain how the third  assumption of the formalism of QMPS issues from the master equation.
The master equation of QMPS in terms of the Hamiltonian operator takes the following form:
\begin{equation*}\label{eq:137}
\hat{H} \psi=i \hbar \frac{\partial \psi}{\partial t}~. 
\end{equation*}
The application of the Hamiltonian operator to a wave function, $\psi=\psi_0 \cdot e^{\frac{i}{\hbar} S}$
yields the following expression:
\begin{equation}\label{eq:139}
\hat{H}\psi=\left(H + \frac{1}{\psi_0}\hat{H} \psi_0 \right) \psi~.
\end{equation}
where $\psi_0$ is a real-valued function.
For convenience, we shall introduce the following complex-valued quantity: 
\begin{equation*}\label{eq:a139}
\tilde{H} = H + \frac{1}{\psi_0}\hat{H} \psi_0~,
\end{equation*}
and refer to it as the function on observable of the Hamiltonian.

Accordingly, the wave equation with respect to the function on observable of the Hamiltonian can be represented as 
\begin{equation}\label{eq:140}
\tilde{H} \psi =i \hbar \frac{\partial \psi}{\partial t}~.
\end{equation}
Dividing both sides of Eq. \ref{eq:139} by $\psi$ and arranging it yields 
\begin{equation*}\label{eq:141}
\tilde{H}=i \hbar \frac{1}{\psi} \frac{\partial \psi}{\partial t}~,
\end{equation*}
\begin{equation}\label{eq:142}
\tilde{H}=i \hbar \frac{\partial \ln \psi}{\partial t}~.
\end{equation}
Now, we introduce $ \tilde{S} $ satisfying the following relation with a wave function, $\psi$:
\begin{equation*}\label{eq:143}
-i \hbar \ln \left\{ \psi \left( \mathbi{q},\mathbi{p},t \right) \right\} = \tilde{S}~,
\end{equation*}
where $ \tilde{S} $ is a complex-valued function dependent on positions, momenta and time.
Arranging the above equation gives
\begin{equation*}\label{eq:144}
\ln{\psi}=\frac{i \tilde{S}}{\hbar} ~.
\end{equation*}
Putting $\tilde{S}=S+iS_0$, we have
\begin{equation*}\label{eq:145}
\psi=\exp \left( -\frac{S_0}{\hbar} \right) \exp \left( \frac{i S}{\hbar} \right)=\psi_0\: \exp \left( \frac{i S}{\hbar} \right)~, 
\end{equation*}
where  $ S_o $ and $ S $ are real-valued.

Consequently, Eq. \ref{eq:142} can be rewritten as
\begin{equation*}\label{eq:146}
\tilde{H}=- \frac{\partial \tilde{S}}{\partial t}~.
\end{equation*}

The real part of the above equation written as
\begin{equation}\label{eq:147}
H+\frac{\partial S}{\partial t}=0
\end{equation}
is nothing but the Hamilton-Jacobi equation.
Then the action, $S$, should be represented as
\begin{equation*}\label{eq:148}
S=\int L \left( \mathbi{q}, \mathbi{p},t \right) \mathrm{d}t~.
\end{equation*}

In this way, we have accounted for the reason why wave functions should have a definite form related to the dynamical quantity of particles, the very action.
Obviously, this outcome results from the requirement that the wave equation should describe not only the probability wave but also mechanical relation of particles.
Therefore the third assumption of this formalism is identical to the requirement that the wave equation should describe  both wave and particle

The core of dissenting arguments about quantum mechanics in phase space lies in what the utility and advantages of this formalism are.
The answer to it is that phase space contains configuration space, and therefore quantum mechanics in phase space amounts to general formalism.
Moreover, mechanics in phase space gives the possibility of delving in depth into the essence of quantum mechanics.
The relation between quantum mechanics in configuration space and that in phase space should be considered to be similar to the relation between the Lagrangian and Hamiltonian formalism in classical mechanics.
In fact, were it not for the Hamiltonian formalism of mechanics, it would not be possible to develop coherent and complete presentation of classical mechanics of many-particle system.
If it is possible to use phase space for quantum mechanics, surely quantum mechanics in phase space acquires the status within quantum mechanics which corresponds to the Hamiltonian formalism of classical mechanics.

Indeed many physicists appear even to doubt the very existence of a true 'reality' at quantum scales and, instead, rely merely upon the quantum-mechanical mathematical formalism to obtain answers to conceptual problems of quantum mechanics \cite{Penrose}.
However, despite all this, it is very remarkable that the Hamiltonian procedures provide the essential background to quantum-mechanical theory.
The present phase-space formalism of quantum mechanics has potential for applying the symplectic geometry to quantum mechanics to describe the probabilistic behavior of microparticle.

\begin{figure}[!th]
\centering
\includegraphics[clip=true, scale=0.45]{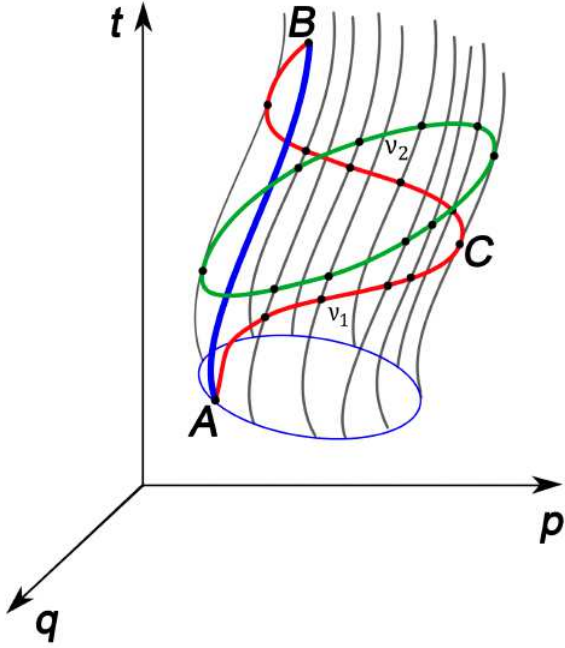}
\caption{\label{fig1}Classical trajectories (black lines) and vortex lines in extended phase space: Ensemble of classical trajectories produces a quantum trajectory represented by the phase flow(red line). The green line, $\nu_2$ denotes the vortex line traversing the ensemble of classical trajectories. The red line and blue line constitute a vortex line, too.}
\end{figure}
Fig.\ref{fig1} shows a family of classical trajectories (black line) belonging to a statistical ensemble in the extended phase space and the vortex lines (green line) traversing them.
Then the classical trajectories constitute a tube in the extended phase space.
On the other hand, the vortex lines represent the phase flow which entails the transition between the classical trajectories.
In microworld, the wave field affects a particle so that it deviates from a definite classical trajectory.
As a result, the particle moves along the fluctuating trajectory which covers a ensemble of classical trajectories.
In this case, the fluctuating trajectory assumes a vortex line.
Then the integrals of the form $\mathbi{p}\mathrm{d}\mathbi{q}-H\mathrm{d}t$ along vortex lines are the same, since they are the Poincar\'{e}-Cartan integral invariant.
In particular, the red line and blue one form a meaningful vortex line.
The red line expresses the positively temporal process and the blue line does the negatively temporal one.
Hereafter, we shall refer to this trajectory expressed by red line enveloping the ensemble of the classical trajectories as the phase trajectory.
It is obvious that this phase trajectory amounts to a quantum trajectory cycling the ensemble of classical trajectories.
The invariance of the Poincar\'{e}-Cartan integral along the vortex line on the same tube of trajectories enables us to quantify a quantum trajectories expressed by this vortex line.
In this case, the Poincar\'{e}-Cartan integral is written as

\begin{flalign}\label{eq:148-1}
&\oint_{\nu_1}\left(\mathbi{p}\mathrm{d}\mathbi{q}-H\mathrm{d}t\right)\nonumber\\
&=\int_{A\rightarrow C \rightarrow B }\left(\mathbi{p}\mathrm{d}\mathbi{q}-H\mathrm{d}t\right)+\int_{B\rightarrow A }\left(\mathbi{p}\mathrm{d}\mathbi{q}-H\mathrm{d}t\right)=h,
\end{flalign}
where $h$ is the constant associated only with the ensemble of classical trajectories.
\begin{figure}[!th]
\centering
\includegraphics[clip=true, scale=0.45]{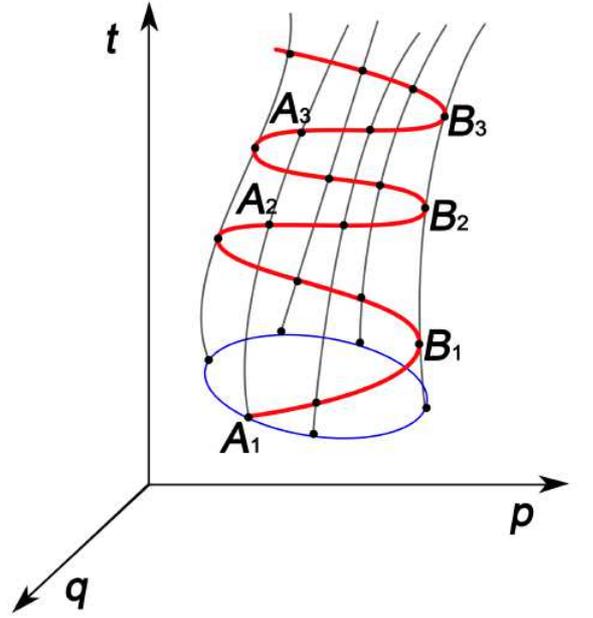}
\caption{\label{fig2}Ensemble of classical trajectories (black lines) and two phase cycles (red lines: $A_1\rightarrow B_1 \rightarrow A_2$ and $A_2\rightarrow B_2 \rightarrow A_3$) }
\end{figure}
The phase trajectory $A\rightarrow C \rightarrow B$ covers the ensemble of trajectories, while $B\rightarrow A$ is a single classical trajectory which can be determined previously with the help of classical mechanics.
From Eq. \ref{eq:148-1} we get
\begin{equation}\label{eq:148-2}
\int_{A\rightarrow C \rightarrow B }\left(\mathbi{p}\mathrm{d}\mathbi{q}-H\mathrm{d}t\right)=h+\int_{A\rightarrow B }\left(\mathbi{p}\mathrm{d}\mathbi{q}-H\mathrm{d}t\right).
\end{equation}
Eq. \ref{eq:148-2} shows that for a given classical trajectory as a part of vortex, the integral of $\mathbi{p}\mathrm{d}\mathbi{q}-H\mathrm{d}t$ along an arbitrary phase trajectories is constant.

For further advance, we examine Fig.\ref{fig2}, where
$A_1\rightarrow A_2 \rightarrow A_3$ expresses a classical trajectory, and $A_1\rightarrow B_1 \rightarrow A_2$ and $A_2\rightarrow B_2 \rightarrow A_3$ are phase trajectories crossing the classical trajectories, respectively.
Namely, a piece of a given classical trajectory corresponds to two phase trajectories succeeding temporally.
Therefore this phase flow represents a temporal evolution of the ensemble of trajectories.
In this case, the sum of the Poincar\'{e}-Cartan integrals for the two vortices is represented as 
\begin{equation}\label{eq:148-3}
\int_{\gamma}\left(\mathbi{p}\mathrm{d}\mathbi{q}-H\mathrm{d}t\right)=2h+\int_{A_1\rightarrow A_3 }\left(\mathbi{p}\mathrm{d}\mathbi{q}-H\mathrm{d}t\right),
\end{equation}
where $\gamma_p$ expresses the phase trajectory, $A_1\rightarrow B_1\rightarrow A_2 \rightarrow B_2 \rightarrow A_3$.
Generally, for $\mathit{N}$ phase trajectories the following relation holds:
\begin{equation}\label{eq:148-4}
\int_{\gamma_p }\left(\mathbi{p}\mathrm{d}\mathbi{q}-H\mathrm{d}t\right)=\mathit{N}h+\int_{\gamma_c }\left(\mathbi{p}\mathrm{d}\mathbi{q}-H\mathrm{d}t\right),
\end{equation}
where $\gamma_p$ denotes the phase trajectory and $\gamma_c$, the classical trajectory.
For a given system, $\int_{\gamma_c }\left(\mathbi{p}\mathrm{d}\mathbi{q}-H\mathrm{d}t\right)$ is definitely determined.
Therefore $\int_{\gamma_p }\left(\mathbi{p}\mathrm{d}\mathbi{q}-H\mathrm{d}t\right)$ represents the phase cycles of trajectory ensemble.
The cyclic property of phase-trajectory integral does not depend on the choice of a particular classical trajectory.
In fact, a given classical trajectory only determines the initial phase of the ensemble of trajectories.
Thus, the action functional can be considered as the phase of quntum-mechanical system.
If $h$ is accepted as the universal constant indicative of the quantum of phase space for arbitrary ensemble of trajectories, then it is nothing but the Plank constant.
Then the quantum phase is represented as
\begin{align}\label{eq:148-5}
2\pi\frac{\int_{\gamma_p }\left(\mathbi{p}\mathrm{d}\mathbi{q}-H\mathrm{d}t\right)}{h}
=\frac{\int_{\gamma_p }\left(\mathbi{p}\mathrm{d}\mathbi{q}-H\mathrm{d}t\right)}{\hbar}=\int_{\gamma_p }\left(\mathbi{k}\,\mathrm{d}\mathbi{q}-\omega\mathrm{d}t\right).
\end{align} 
Evidently. the above expression indicative of the de Broglie relation implies that the existence of $h$ gives rise to the space-time quantization.
In the end, we reach the conclusion that the wave function should be represented as a function with the phase part,
\begin{equation}\label{eq:148-5}
\exp\left\{i\frac{\int_{\gamma_p }\left(\mathbi{p}\mathrm{d}\mathbi{q}-H\mathrm{d}t\right)}{\hbar}\right\}=\exp\left\{i\frac{S}{\hbar}\right\}.
\end{equation} 

This is the explanation for the $\mathbf{assumption\, III}$  from the viewpoint of symplectic geometry, which makes the assumptions of this formalism reasonable and unifies them.

In conclusion, this perspective on quantum dynamics brings to a successful conclusion that helps to resolve the conceptual problems of quantum mechanics.

The essentials are summarized as follows.

$\mathbf{a:\,}$ \textit{The quantum of action, h gives rise to the ensemble of discontinuous classical trajectories.}

$\mathbf{b:\,}$ \textit{The wave field causes the cycling of phase trajectories.}
This means the probabilistic occupation of classical trajectories and entanglement of the trajectories.
The non-locality is due to it.

$\mathbf{c:\,}$ \textit{The density of the phase points is determined by the phase trajectories traversing the ensemble of discontinuous classical trajectories}.
Then the density of the phase points gives the probability density.

$\mathbf{d:\,}$ \textit{A particular event depends on a definite phase of phase flow}. 
On the other hand, the phase is not deterministic at the moment of measurement, and thus the results are to obey the probabilistic law associated with the phase trajectories.
We consider that this view gives the key to the mystery of quantum mechanics involving the problems with the locality, determinism, measurement and otherwise.

Thus, this formalism whittles down the significance of the assumptions considerably and provides the possibility of resolving the conceptual and philosophical problems with a natural epistemology.

\section{\label{sec:Summary}Summary and discussion}

We summarize the main results.

First, we have obtained an alternative master equation of QMPS, which may newly interpret the foundation of quantum mechanics and may help to formulate a more generalized theory of quantum mechanics. 
Our research shows that the wave equation for microparticles can be readily obtained without a jump of logic, provided that the wave field is regarded as corresponding to a statistical ensemble in phase space. 
It is important to note that the  conceived equation is derived independently of the Schr\"{o}dinger equation. 
On the other hand, this equation yields the schr\"{o}dinger equation by admitting a definite approximation.
For further development of quantum mechanics, it is desirable to establish more natural, more essential epistemology. 
Our purpose is to show that such a clue can be found from the consideration of the statistical ensemble in phase space embodying the wave field as a physical reality. 

Second, we have shown how the master equation of QMPS can be applied to some simple cases to obtain reasonable results of quantization.
The obtained results illustrate that the view is understandable that in some sense, quantum mechanics becomes a part of statistical mechanics \cite{Wyatt,Holevo}.

Third, we have interpreted approximation of the formalism of QMPS to the Schr\"{o}dinger equation and probabilistic and mechanical structure of the present formalism. 
For this formalism, it is indispensable to introduce the notion of operators inasmuch as with the help of the operators, the relations of this formalism with the other formulations of quantum mechanics are elucidated and the wave-particle constitution of quantum mechanics is to be revealed. 
With the aid of  operators, this formalism explains how the master equation of QMPS approximates to the Schr\"{o}dinger equation, and how it contains the mechanical relation of particles.
Especially, QMPS is the formalism capable of using at once the phase-space variables and the corresponding operators. 

Fourth, we have shown that this formalism provides the possibility of resolving the relativistic problems. 
It is possible to obtain the relativistic wave equation by simply inserting the relativistic phase velocity into the master equation. 
Therefore this formalism shuns the hardships faced by the relativistic quantum theory as in Dirac and Klein-Gordon's equations.
As to form of equation, the master equation of QMPS is identical with Dirac's equation in that it is represented as a first-order partial differential equation. 
Such characteristics have us avoid the serious problems concerning the negative energy in the relativistic case, and offer convenience for mathematical treatment.

The master equation is distinguished from other kinds of the causal theory of quantum mechanics since it is  directly obtained by considering the statistical ensemble in phase space, without recourse to the Schr\"{o}dinger equation. 

We summarize the main differences of this formalism from the others.

First, unlike the phase-space formalisms using the Wigner function and the Weyl map, this formalism is not a theory dealing with the map of the wave functions in configuration space to the probability density in phase space.
The phase-space formalism using the Wigner functions and the Weyl correspondence merely addresses the transformation of representation between phase space and Hilbert space. Evidently, the phase-space formalism clarifies the equivalence of phase-space quantization to the conventional formulation of quantum mechanics in Hilbert space. In essence, it is a theory studying  the mapping of the wave function in configuration space to phase-space function.
In this respect, the present formalism is also different from the tomography theory.
The theory of quantum tomography leads the Hilbert-space quantum mechanics to the phase-space quantum mechanics by treating the mapping of quantum states in position space to ones in phase space. This theory is also a kind of the mapping theories relevant to quantum states. Without adopting the Schr\"{o}dinger equation and its solution, the tomography theory cannot work.
On the contrary, our formalism does not address mapping. Instead of map,  it determines the self-reliance wave function by means of self-standing equation of quantum mechanics in phase space.
Therefore the differences are due largely to the autonomy of formalism.

Second, unlike the Hilbert-space formalism, this formalism starts with the wave function and probability density in phase space, which represent the wave field considered as a physical reality. The wave field is considered to produce the statistical ensemble for microparticles.
The main difference is that the two formulations use different spaces.

In spite of all the differences, this formalism is closely related to the causal theories of quantum mechanics, since the master equation is described in phase space which entails the trajectory with a definite probability. 
On the other hand, our formalism can be regarded as forming a bridge between the standard theory of quantum mechanics and the causal theories of quantum mechanics. 
This is because our formalism is transformed into quantum mechanics in configuration space by integrating over phase space under definite assumptions, and provides detailed accounts of why the operators should be introduced into quantum mechanics. 

Within the framework of this formalism, we have obtained an alternative master equation for microparticles, by beginning with statistical ensemble in phase space due to the wave field. 
The wave equation involves both the probability continuity equation and the energy relation for particles. 
Essentially, the probability continuity equation is a quantal version of Liouville's equation taking into account the quantal correlation between trajectories via the wave field represented in phase space. 
This correlation exhibits the wave-like property of microparticles. 

This formalism uses the potential operator instead of the potential function. 
It can be regarded as a novel, reasonable result obtained by this formalism. 
In fact, since the classical potential is to be determined by force and path, it is impossible to take the classical potential for the quantal potential without admitting some approximation. 
As a classical path loses its meaning in the quantum world, also so does  the classical potential.
The master equation of QMPS evidently explains this matter. 
This formalism adopts the potential in the statistical way, thus reproducing the virial theorem in quantum mechanics.
It is possible to explain with the aid of this formalism what approximation the Schr\"{o}dinger equation makes. 
According to the interpretation of this formalism, a definite probability corresponds to a given trajectory. 
In connection with this fact, it is reasonable to consider the master equation of QMPS as the quantal version of Liouville's equation.

It is necessary to review the uncertainty relation.
The below argument, in fact, is the summarized citation of Home and Whitaker's description \cite{Home}.

It is interesting to note that the derivation of the uncertainty principle uses no input from quantum dynamics. 
Even if one uses a wave-function having the wrong symmetry and violating the Schr\"{o}dinger equation, the uncertainty relation will not necessarily be violated. 
The uncertainty principle is thus insensitive to any modification of the Schr\"{o}dinger equation.
The interpretational significance of the uncertainty principle may be stated in one of the two following ways. 
The first corresponds to Heisenberg's gedanken experiments. 
This approach to the Heisenberg principle has been quite widely held right up to the present day. 
The second approach to interpreting the uncertainty principle  recognizes that uncertainty in the value of a dynamical variable refers to the statistical spread over the measured values for the various identical members of the ensemble of systems. 
The operational significance of the two approaches is totally different. 
In the second, there is no question of simultaneous measurement of the dynamical variables related to a single particle, while this concept forms the essence of the first approach, where the uncertainty in a single measurement is interpreted as the estimate of imprecision in the measured value of a dynamical variable for a single particle. 
It is obvious that Heisenberg' s thought experiments in the original form should certainly not be regarded as providing a proof of the uncertainty principle.
In fact, the uncertainty relation is a natural consequence quantum mechanics yields \cite{Home}. 

Our approach can adduce adequate reasons in support of the second approach.
The master equation is the quantal version of Liouville's equation, and therefore the uncertainty principle becomes an inevitable corollary of the present equation.
In this connection, what is most important for quantization is the existence of the quantum of phase space, i.e. $\hbar$ rather  than the uncertainty relation adopted by Heisenberg.
In fact, the Schr\"{o}dinger equation does not involve the first approach to interpreting the uncertainty principle.
Should we not introduce the one-to-one correspondence between position and momentum according to the scheme, \ref{eq:sp1}, we would not obtain the Schr\"{o}dinger equation.   
All these arguments conclude that the uncertainty relation embodies the statistical spread of ensemble in phase space which is ruled by quantum laws.

It is necessary to review the problem of whether the Schr\"{o}dinger equation is mathematically rigorous.
Starting from the definition of mean value, we can explicitly demonstrate that the Schr\"{o}dinger equation cannot avoid some approximations besides non-relativistic one.
For convenience, we consider the Schr\"{o}dinger equation for one particle.
By definition, the mean value of momentum component $p_x$ reads
\begin{equation}\label{eq:149}
\left<p_x\right >=\int \psi^{*}\hat{p}_x \psi \mathrm{d} v
=\int \psi^{*}\left(\frac{\hat{p}_x \psi  }{\psi}\right)~\psi \mathrm{d} v .
\end{equation}
Obviously, the real part of $\dfrac{\hat{p}_x \psi  }{\psi} $ is $x-$component of momentum.
Therefore we can write $\dfrac{\hat{p}_x \psi  }{\psi} $ as

\begin{equation}\label{eq:150}
\frac{\hat{p}_x \psi  }{\psi}=\tilde{p_x}=p_{x-real} +i p_{x-imag},
\end{equation}
where $p_{x-real}$ and $p_{x-imag}$ are the real and imaginary part of $\dfrac{\hat{p}_x \psi  }{\psi} $, respectively.
Generally, $\tilde{p_x}=\dfrac{\hat{p}_x \psi  }{\psi}$ is a complex function dependent on coordinates, since $\psi$ is not the eigen function of $\hat{p}_x $.
As a result, we easily arrive at 
\begin{align}\label{eq:150}
\hat{p}_x^2 \psi  =\hat{p}_x \hat{p}_x  \psi =\hat{p}_x \left( \tilde{p_x} \psi\right )=\tilde{p_x}\hat{p}_x \psi+\psi \hat{p}_x \tilde{p_x} \nonumber \\
=\tilde{p_x}^2 \psi+\psi \hat{p}_x \tilde{p_x} \neq \left( \mathrm{Re} \tilde{p_x} \right )^2\psi=p_{x-real}^2 \psi ,
\end{align}
where $\mathrm{Re}$ denotes the real part of complex number.
On the other hand, the Schr\"{o}dinger equation is obtained in terms of the energy relation
\begin{equation}\label{eq:151}
E=\frac{\mathbi{p}^2}{2m}+U .
\end{equation}
The operator corresponding to Eq. \ref{eq:151} reads
\begin{equation}\label{eq:152}
\hat{E}=\frac{\hat{\mathbi{p}}^2}{2m}+U .
\end{equation}
Consequently, the wave equation for this operator is written as
\begin{equation}\label{eq:153}
\hat{E}\psi=\left(\frac{\hat{\mathbi{p}}^2}{2m}+U\right) \psi .
\end{equation}
It is this equation that is the Schr\"{o}dinger equation.
According to Eq. \ref{eq:150}, it is well-grounded that Eq. \ref{eq:153}, i. e. the Schr\"{o}dinger equation is assessed as neglecting some terms.
It is obvious that only when these terms are negligible, the Schr\"{o}dinger equation gives reasonable solutions.
Consequently, the requirement for approximation to the Schr\"{o}dinger equation is that $\tilde{p_x}$ can approximate to a real constant.
From the above argument, it follows that the double application of differential operator such as the momentum operator to wave function violates the exact correspondence relation between the operator and dynamical quantity.
Especially, for the case of real-valued wave function we encounter an intractable problem.
In this case, $\dfrac{\hat{p}_x \psi  }{\psi} $ via a single application of $\hat{p}_x $ to the wave equation becomes a purely imaginary number. 
This indicates that the momentum vanishes.
On the other hand, the calculation of $\dfrac{1}{2m}\dfrac{\hat{p}_x^2\psi  }{\psi} $ via a double application of $\hat{p}_x $ to the wave function gives a purely real number which means nonzero kinetic energy.
This result shows that despite zero momentum the corresponding kinetic energy may have a nonzero value.
It is this fact that demonstrates the approximate aspect of Schr\"{o}dinger equation. 
Such a situation significantly emphasizes the necessity of accepting the phase-space formalism not involving the abovementioned inconsistency.

The present theory is based on phase space, and therefore the concept of trajectory is naturally accepted.
In general, the purpose of quantum trajectory theory is to in-depth understand the nature of quantum mechanics and to provide classical-like insight into the dynamics and
physics of the quantum processes \cite{Poirier}.
However, it should be emphasized that this theory is not a return to classical mechanics, since it uses the wave function which determines the quantum trajectory distinguished from the classical one, reflecting the quantum nature.
The present formalism is expected to provide a new possibility of  shedding light on the foundation of quantum mechanics in terms of a theory free of paradoxes and to promote an understanding as clear as that of classical mechanics, as it is a version of the quantum trajectory theory orienting itself to a realistic description
of microparticles' motion by virtue of primordial intuition about mechanics.
In fact, it is natural for this phase-space formalism to introduce the interpretation achieved by the quantum trajectory theory, since it is identified with other quantum trajectory theories in the sense of trajectory \cite{Faraggi,Floyd,Bittner2,Holland2}. 

We still have not succeeded in obtaining the analytical solution to the problem of hydrogen atom within the framework of the present theory.
It may be that it is impossible to get the desired analytical solution.
However, the failure does not devalue the present theory because while it involves the Schr\"{o}dinger equation as its special case, the Schr\"{o}dinger equation gives the analytical solution to the problem of hydrogen atom.

\section{Conclusion}\label{concl}

In this work, we have investigated an alternative formalism of quantization in terms of statistical ensemble in phase space, keeping the statistical perspective on quantum mechanics. 

Our work has shown that it is possible to establish an alternative autonomous formalism of quantum mechanics in phase space, starting from statistical ensemble in phase space. 
This formalism produces within its framework the master equation without recourse to the other formulations of quantum mechanics. 
Manipulating within the framework of its theory, this formalism provides  a series of the calculations and interpretations of quantization in phase space. 
With the help of the master equation of this formalism have been explained the structure of quantum mechanics in phase space and the approximation to the Schr\"{o}dinger equation. 

Up to now, the standard formulation of quantum mechanics has been certainly the most successful, so it still keeps its dominant position in developing the science for microworld, pursuing essentially different picture from that of classical mechanics. 
Nevertheless, there exist different formulations of quantum mechanics including QMPS. 
This fact means that individual formalism of quantum mechanics has its inherent merits irreplaceable by the others.
Of course, the present formalism of quantum mechanics may not be superior to the standard theory of quantum mechanics in some problems of calculation. 
However this fact does not lead the argument to indicating that this formalism is inferior to the standard formalism. 
To understand such a context, it is enough to recall that the present formalism includes the configuration-space formalism as its special case.
The two formulations distinguish themselves regarding whether to use phase space or position space.
It should be noted that if the phase-space formalism is possible, it  become a generalized theory comprising the configuration-space formalism.
The present formalism brings benefits to revealing with a natural epistemology the nature of quantum mechanics and the relations between several formulations of quantum mechanics.

Especially, it is expected that the present phase-space formalism provides the potential to resolve open problems including the relativistic quantum theory and the connection between classical and quantum mechanics, and to lay the  foundation for constructing quantum mechanics in phase space as an autonomous formalism. 
Obviously, the relativistic wave equation of the present theory is a generalized equation which yields the non-relativistic equation within the limit of $c \gg v$ .
It is straightforward that since wave functions are defined in phase space, this formalism is able to offer a new possibility of improving mathematical foundations  and  interpretations of quantum mechanics.

In conclusion, our work confirms that the formalism of quantization in terms of statistical ensemble in phase space is consistent with the fundamentals of quantum mechanics, and offers a possibility of resolving some intractable problems arising from other formulations of quantum mechanics.
  
\section*{Acknowledgments}
This work was supported partially from the Committee of Education, Democratic People's Republic of Korea, under the project entitled ``Statistical Formalism of Quantum Mechanics''.
 We thank Profs. Chol-Jun Yu and Hak-Chol Pak from Kim Il Sung University and Profs. Il-Hwan Kim and Se-Hun Ryang from the University of Science for useful advice and help. 
Prof. Nam-Hyok Kim from Kim Il Sung University and Prof. Yon-Il Kim from  the State Academy, DPR Korea are appreciated for valuable discussion. 
Authors would like to thank the editor and anonymous reviewers for comment and advice.

\section*{Conflict of interest}
The authors declare no competing financial interest.



\end{document}